\documentclass[a4paper,fleqn,usenatbib,useAMS]{mnras}

\usepackage{hyperref}
\newcommand{\RNum}[1]{\uppercase\expandafter{\romannumeral #1\relax}}
\usepackage{caption,booktabs}

\usepackage{graphicx}
\usepackage{amsmath}
\usepackage{amssymb}
\usepackage{multicol}
\usepackage{bm}
\usepackage{pdflscape}

\usepackage[T1]{fontenc}
\usepackage{ae,aecompl}
\usepackage{newtxtext,newtxmath}

\title[CHILES \RNum{6}]{CHILES \RNum{6}: HI and H$\alpha$ Observations for z $<$ 0.1 Galaxies; \;\;\;\;\;\;\;\;\;\;\;\;\;\;\;Probing HI Spin Alignment with Filaments in the Cosmic Web}

\author[Blue Bird et al.]{
J. Blue Bird$^{1}$\thanks{julia@astro.columbia.edu},
J. Davis$^{2}$\thanks{jdavis@astro.wisc.edu},
N. Luber$^{3,4}$ ,
J.H. van Gorkom$^{1}$,
E. Wilcots$^{2}$,
D.J. Pisano$^{3,4,5}$, \newauthor
H. B. Gim$^{6,7}$,
E. Momjian$^{8}$,
X. Fernandez$^{9}$,
K. M. Hess$^{10,11}$,
D. Lucero$^{12}$,
R. Dodson$^{13}$, \newauthor
K. Vinsen$^{13}$,
A. Popping$^{13,14}$,
A. Chung$^{15}$,
K. Kreckel$^{16}$,
J. M. van der Hulst$^{11}$,
and M. Yun$^{6}$\\
\\
$^{1}$Department of Astronomy, Columbia University, 550 West 120th Street, New York, NY 10027, USA\\
$^{2}$Department of Astronomy, University of Wisconsin - Madison, 475 N Charter St., Madison, WI 53706, USA\\
$^{3}$Department of Physics and Astronomy, West Virginia University, P.O. Box 6315, Morgantown, WV 26506, USA\\
$^{4}$Center for Gravitational Waves and Cosmology, West Virginia University, Chestnut Ridge Research Building, Morgantown, WV 26505\\
$^{5}$Adjunct Astronomer at Green Bank Observatory, Green Bank, WV, USA\\
$^{6}$Department of Astronomy, University of Massachusetts, Amherst, MA 01003, USA\\
$^{7}$School of Earth and Space Exploration, Arizona State University, 781 Terrace Mall, Tempe, AZ 85287 USA\\
$^{8}$National Radio Astronomy Observatory, P.O. Box 0, Socorro, NM 87801, USA\\
$^{9}$Department of Physics and Astronomy, Rutgers, The State University of New Jersey, Piscataway, NJ 08854-8019, USA\\
$^{10}$ASTRON, the Netherlands Institute for Radio Astronomy, Postbus 2, Dwingeloo NL-7900AA, The Netherlands\\
$^{11}$Kapteyn Astronomical Institute, University of Groningen, Landleven 12, 9747 AD, Groningen, The Netherlands\\
$^{12}$Department of Physics, Virginia Tech, 850 West Campus Drive, Blacksburg, VA 24061, USA\\
$^{13}$International Centre for Radio Astronomy Research, The University of Western Australia, Crawley, WA 6009, Australia\\
$^{14}$Australian Research Council, Centre of Excellence for All-sky Astrophysics (CAASTRO), Australia\\
$^{15}$Department of Astronomy, Yonsei University, 50 Yonsei-ro, Seodaemun-gu, Seoul 03722, Korea\\
$^{16}$Max Planck Institute for Astronomy, Knigstuhl 17, D-69117 Heidelberg, Germany
}

\date{Accepted 2019 November 27. Received 2019 November 27; in original form 2019 September 11}
\pubyear{2019}

\begin{document}
\label{firstpage}
\pagerange{\pageref{firstpage}--\pageref{lastpage}}
\maketitle

\begin{abstract}
We present neutral hydrogen (HI) and ionized hydrogen (H$\alpha$) observations of ten galaxies out to a redshift of 0.1. The HI observations are from the first epoch (178 hours) of the COSMOS HI Large Extragalactic Survey (CHILES). Our sample is HI biased and consists of ten late-type galaxies with HI masses that range from $1.8\times10^{7}$ M$_{\odot}$ to $1.1\times10^{10}$ M$_{\odot}$. We find that although the majority of galaxies show irregularities in the morphology and kinematics, they generally follow the scaling relations found in larger samples. We find that the HI and H$\alpha$ velocities reach the flat part of the rotation curve. We identify the large-scale structure in the nearby CHILES volume using DisPerSE with the spectroscopic catalog from SDSS. We explore the gaseous properties of the galaxies as a function of location in the cosmic web. We also compare the angular momentum vector (spin) of the galaxies to the orientation of the nearest cosmic web filament. Our results show that galaxy spins tend to be aligned with cosmic web filaments and show a hint of a transition mass associated with the spin angle alignment.
\end{abstract}


\section{Introduction}
In the last decade, tremendous progress has been made in our understanding of structure formation in the Universe. Simulations of dark matter show that structures develop over time forming the large scale structure of the Universe -- the so-called cosmic web consisting of walls, filaments, and voids. Observations show that galaxies lie in this interconnected cosmic web and that stars trace the dark matter. However, how galaxies grow and evolve is less well understood. 

Ongoing infall of gas from the intergalactic medium is thought to be important for the growth of galaxies, with cold gas accretion from the cosmic web likely a dominant factor \citep{SanchezAlmeida2014}. Simulations and models highlight physical processes that may be relevant in studying large scale environments. Theorists distinguish two modes of gas infall: 'hot mode' accretion where gas gets shock heated to high temperatures and then slowly cools and settles in a disk, and 'cold mode' accretion where much cooler gas flows along filaments directly into the disk. 'Cold mode' accretion is a favorable scenario for gas accretion onto galaxies and simulations make very specific predictions about its dependence on galaxy mass, environment, and redshift \citep{Keres2005}. Recently, \cite{Aragon-Calvo2019} proposed the halting of cold gas accretion through cosmic web detachment (CWD). In this picture, galaxies accrete cold gas from the cosmic web until they enter regions of crossing velocity streams, including regions near large filament backbones. These interactions detach galaxies from the cosmic web and sever their cold gas supply, quenching star formation. As the interaction between galaxies and the neutral hydrogen (HI) reservoir in filaments is poorly constrained, more work is needed to confidently measure the influence large-scale structure has on galaxy evolution.

There has been a growing number of methods to quantify large-scale environment, \cite{Libeskind2018} provide a review and a comparison of different methods. Methods such as Discrete Persistent Structure Extractor (DisPerSE) \citep{Sousbie2011}, a scale-free topological algorithm that uses Morse theory and Delaunay tessellations, are used to characterize the large-scale distribution of galaxies. It is by now well established that galaxy properties, such as stellar mass, color-type and star formation rate (SFR), show dependence on location in large-scale environment \citep{Chen2017, Malavasi2017, Kuutma2017, Laigle2018, Kraljic2018, Luber2019}. More recently, HI as a function of distance to cosmic web filaments has been investigated with conflicting results \citep{Kleiner2017, CroneOdekon2018}. In addition to location, there are predictions about the orientation of galaxies with respect to large-scale structure \citep{Porciani2002, Codis2015}. A key prediction is that low-mass galaxies tend to align their angular momentum vector (spin) with their nearby filament while high-mass galaxies tend to have their spin perpendicular to their nearby filament. This is seen in both dark matter simulations for halos \citep{Wang2018, GaneshaiahVeena2018} and hydrodynamic simulations for galaxies \citep{Codis2018, Kraljic2019}. Observational studies of galaxy spin-filament alignments show mixed results. Hints of spin alignment for galaxies have been identified by \citet{Tempel2013, Tempe&Libeskind2013, Pahwa2016, Chen2019, Welker2019} while \citet{Krolewski2019} find no evidence for alignment. The next generation of HI surveys, such as the COSMOS HI Large Extragalactic Survey (CHILES), will provide unique observations to compare to these studies and predictions.

While ongoing infall of gas from the intergalactic medium is thought to be important for the growth of galaxies, observational evidence of accretion remains challenging to obtain. Due to the intrinsic faintness of HI 21-cm emission, it has been difficult to probe beyond a redshift of z $\sim$ 0.1 without prohibitively long integration times. Large single-dish radio surveys such as HIPASS \citep{Barnes2001} and ALFALFA \citep{Giovanelli2005} have compiled a large number of HI-detected galaxies, but only to a redshift of z $\sim$ 0.06, and with relatively low angular resolution. At higher resolution, targeted interferometric surveys of varying galaxy type and environment such as WHISP \citep{vanderHulst2001}, THINGS \citep{Walter2008}, VIVA \citep{Chung2009}, HALOGAS \citep{Heald2011}, Little THINGS \citep{Hunter2012}, and VGS \citep{Kreckel2012} have uncovered numerous interesting HI features potentially linked to formation or accretion processes. From these observations (see \citet{Sancisi2008} for a review), we have learned much about HI distributions and kinematics and how we might infer the presence of accretion. Accretion phenomena become evident in the outskirts and extraplanar regions of spiral galaxy disks, requiring deep 21-cm emission investigations to search for lower column density and/or anomalous velocity range gas. With long integration times, the CHILES survey is beginning to probe this regime beyond the most local galaxies.

CHILES is an HI survey using the upgraded Karl G. Jansky Very Large Array (VLA). For the first time, we are imaging the HI distribution and kinematics in a single pointing of continuous redshift range 0 $<$ z $<$ 0.45. CHILES will measure the HI gas reservoir over a substantial look-back time and provide HI content, morphology, and kinematics for a wide range of stellar masses and environments. CHILES will produce HI images of at least 300 galaxies across the entire redshift range, with a linear resolution of 350 pc, 19 kpc, and 42 kpc at z = 0.03, z = 0.20 and z = 0.45 respectively. The survey will be able to detect at the highest redshift $3.0\times10^{10}$ M$_{\odot}$ at 5$\sigma$, assuming a 150 km s$^{-1}$ profile width. The $40'\times40'$ pointing in the COSMOS field \citep{Scoville2007} is chosen such that it has no strong continuum sources. The COSMOS field is ideal for a survey like this because of the wealth of ancillary data.

In this paper, we present results for ten galaxies out to a redshift of 0.1, from the first epoch of the CHILES survey. At 178 hours, this is a unique amount of observing time as nearby galaxies are typically observed for only a few hours resulting in column density sensitivities an order of magnitude lower at similar resolution. We utilize this first epoch of data as a science verification study for the CHILES survey. Our results reveal irregularities in the morphologies and kinematics in most of the sample. Our results show a tendency of galaxy spins to be aligned with cosmic web filaments and possibly the existence of a transition mass where the alignment changes.

In addition to the HI data, we study ionized gas kinematics with optical long-slit data obtained from the Southern African Large Telescope (SALT). Observations of optical emission lines trace population I stars, particularly HII regions associated with star-forming regions in the galactic disk. These lines are a good tracer of the overall circular motion of the disk given that they have small velocity dispersion compared to the rotation velocity. In this paper, we examine the structural relation of HI and H$\alpha$ disks with rotation curves of H$\alpha$ in the inner regions and with HI in the outer regions. The results reveal that the flat part of the rotation curve is reached for the H$\alpha$ data, along with disturbed kinematics due to non-circular motion in the inner region of one of the irregular galaxies in our sample.

This paper is organized as follows. In Section 2, we outline the observations and data reduction for HI and H$\alpha$. In Section 3, we describe the data analysis, sample properties, individual galaxies, and the derivations of the galaxy environments. In Section 4, we analyze the stellar, HI and H$\alpha$ content including HI gas fraction, HI deficiency, HI size-mass relation, HI and H$\alpha$ line width comparison, and HI and H$\alpha$ baryonic Tully-Fisher relation. In Section 5, we discuss the HI morphology and kinematics as well as the HI properties as a function of distance and orientation to the cosmic web. Section 6 covers the conclusion. In the Appendix, we present figures for the individual galaxies.

Throughout this paper, we use J2000 coordinates, velocities in the optical convention, and a barycentric reference frame. This paper adopts a flat $\Lambda$CDM cosmology using H$_{o}$ = 67.3 km s$^{-1}$ Mpc$^{-1}$ and $\Omega_{M}$ = 0.316 \citep{Planck2018} to calculate distances and physical sizes.

\section{Observations}
We carried out a 60-hour pilot study \citep{Fernandez2013} (CHILES \RNum{1}) of the CHILES field during the commissioning of the upgraded VLA correlator. We imaged HI in the redshift range 0 $<$ z $<$ 0.193 and found 33 detections, from which we draw a sample to study with our 178-hour first epoch of data. The full CHILES survey has been underway since late 2013 when observations began on the VLA. In \citet{Dodson2016} (CHILES \RNum{2}), we compared the suitability of different computing environments for processing a data set like CHILES. From the first 178 hours, we already have dozens of HI detections, ranging from the z $<$ 0.1 galaxies covered in this paper (CHILES \RNum{6}), to galaxy groups at z = 0.12 and z = 0.17 in \citet{Hess2019} (CHILES \RNum{4}), to the highest redshift HI detection so far at z = 0.37 in \citet{Fernandez2016} (CHILES \RNum{3}). In a separate paper, \citet{Luber2019} (CHILES \RNum{5}) explore the use of DisPerSE to identify cosmic web filaments in the CHILES volume. For this work, we utilize the complementary multi-band photometry and optical redshift information from the G10/COSMOS v05 catalog \citep{Andrews2017}.

\begin{table}
\caption{CHILES Observation Details for Our Sample}
\begin{tabular}{lc}
\hline
Survey Epoch &Epoch 1\\
Observation Date &2013-2014\\
Array Configuration &VLA-B\\
Integration [hr] &178\\
Bandpass and Flux Density Scale Calibrator &3C286\\
Phase Calibrator &J0943-0819\\
Frequency Coverage (MHz) &1300 - 1411\\
Redshift Range &0.0068 - 0.0930\\
Synthesized Beam (") &$6.4\times4.7$$^{c}$ - $6.8\times5.1$$^{d}$\\
Frequency Resolution (kHz)$^{b}$ &62.5\\
Velocity Resolution (km s$^{-1}$)$^{b}$ &13.3$^{c}$ - 14.4$^{d}$\\
Spatial Resolution (kpc) &0.8$^{c}$ - 10.7$^{d}$\\
RMS Noise ($\mu Jy$ beam$^{-1}$ channel$^{-1}$)$^{b}$ &76.0$^{c}$ - 83.0$^{d}$\\
Typical 1$\sigma$ N$_{HI}$ (cm$^{-2}$ channel$^{-1}$)$^{b}$ &3.3$^{c}$ - 4.4$^{d}\times10^{19}$\\
\hline
\multicolumn{2}{p{8cm}}{\footnotesize{$^{a}$After Hanning smoothing plus additional velocity smoothing. $^{b}$At z = 0.0068. $^{c}$At z = 0.0930.}}
\end{tabular}
\label{table1}
\end{table}

The 1000-hour survey is divided into epochs spread over several years to make all observations in the VLA's B-configuration (see Table 1 for observation properties). Observations of the first epoch completed in 2014, totaling 178 hours. The first epoch is divided into several observing sessions with varying lengths of 1 to 6 hours. A 6-hour session consists of approximately 5 hours on-source and 1 hour for calibration and setup. The 480 MHz frequency span with dual-polarization is tuned to a range of 970 to 1450 MHz, corresponding to a redshift interval of z = 0 to 0.45. Frequency dithering is used with three settings to minimize sensitivity loss at the edge of sub-bands. The setup includes 15 sub-bands of 32 MHz, with each sub-band having 2048 channels of 15.6 kHz (3.3 km s$^{-1}$ at z = 0), which are then Hanning smoothed to 31.2 kHz (6.6 km s$^{-1}$ at z = 0).

Data reduction is carried out in CASA \citep{McMullin2007} using a modified version (1.2.0) of the NRAO continuum pipeline optimized for spectral line data. Each observing block is calibrated separately, with a combination of machine and manual flagging. All of the blocks are then imaged together using CASA and Amazon Web Services (as described in \citet{Dodson2016}). Initial image cubes (with a pixel scale of 2") are $4096\times4096$ pixels to include out of field sources that introduce side lobes through the main field of interest. These sources are modeled for cleaning. The images are made using Briggs weighting with a robustness factor of 0.8 and cleaned with 10,000 iterations to remove the side lobes. The image cubes are then scaled down to $2048\times2048$  pixels which include the full primary beam (z = 0). The continuum is subtracted in the image plane with a first-order polynomial fit. Image sub-cubes of 4 MHz are made consisting of 64 channels averaged to a 62.5 kHz frequency resolution (13.3 km s$^{-1}$ at z = 0). Smaller sub-cubes with individual galaxies are cleaned down to 1$\sigma$ of the RMS using a box around the emission region.

\begin{table}
\caption{SALT Observation Details}
\begin{tabular}{ccccccc}
\hline
C08 &Date &Grating &Slit &V &Exp. &PA$_{H{\alpha}}$\\
ID & & &[''] & &[sec] &[deg]\\
(1) &(2) &(3) &(4) &(5) &(6) &(7)\\
\hline
1213496 &20160501 &PG2300 &1.5 &19.1 &800 &214, 305\\
1180660 &20160501 &PG2300 &1.5 &17.4 &600 &84, 137\\
1197518 &20160501 &PG2300 &1.5 &19.6 &1700 &163, 46\\
1204837 &20170423 &PG2300 &1.25 &17.7 &880 &175\\
1227948 &20160502 &PG2300 &1.5 &19.6 &1700 &54, 290\\
1432731 &20170424 &PG1800 &1.25 &18.8 &1600 &101\\
1437568 &20161229 &PG1800 &1.25 &18.6 &750 &258, 18\\
969633 &20170423 &PG1800 &1.25 &18.6 &1600 &240\\
1419315 &20170131 &PG1800 &1.25 &19.8 &1600 &240, 283\\
1221696 &20161229 &PG1800 &1.25 &19.8 &1600 &180, 153\\
\hline
\multicolumn{7}{p{8cm}}{\footnotesize{(1) COSMOS 08 ID; (2) Observation date; (3) Grating; (4) Slit size; (5) V-band magnitude; (6) Total exposure time per position angle; (7) H${\alpha}$ position angles.}}
\end{tabular}
\label{table2}
\end{table}

In this work, we focus on galaxies in the frequency range of 1300 to 1411 MHz which corresponds to a redshift range of z = 0.0068 to 0.0930. Overall, we achieve a mean RMS of 80 $\mu Jy$ beam$^{-1}$ per 62.5 kHz channel throughout our ten image cubes, which is close to the theoretical noise. The channel resolution corresponds to 13.3 km s$^{-1}$ at z = 0.0068 and 14.3 km s$^{-1}$ at z = 0.0930. The final image resolution is $6.4\times4.7$" and $6.8\times5.1$", in the z = 0.0068 and z = 0.0930 cubes, respectively. We reach 1$\sigma$ column densities of $3.3\times10^{19}$ cm$^{-2}$ (13.3 km s$^{-1}$ channel) at z = 0.0068 and $4.4\times10^{19}$ cm$^{-2}$ (14.3 km s$^{-1}$ channel) at z = 0.0930.

\begin{figure*}
\includegraphics[scale=0.69]{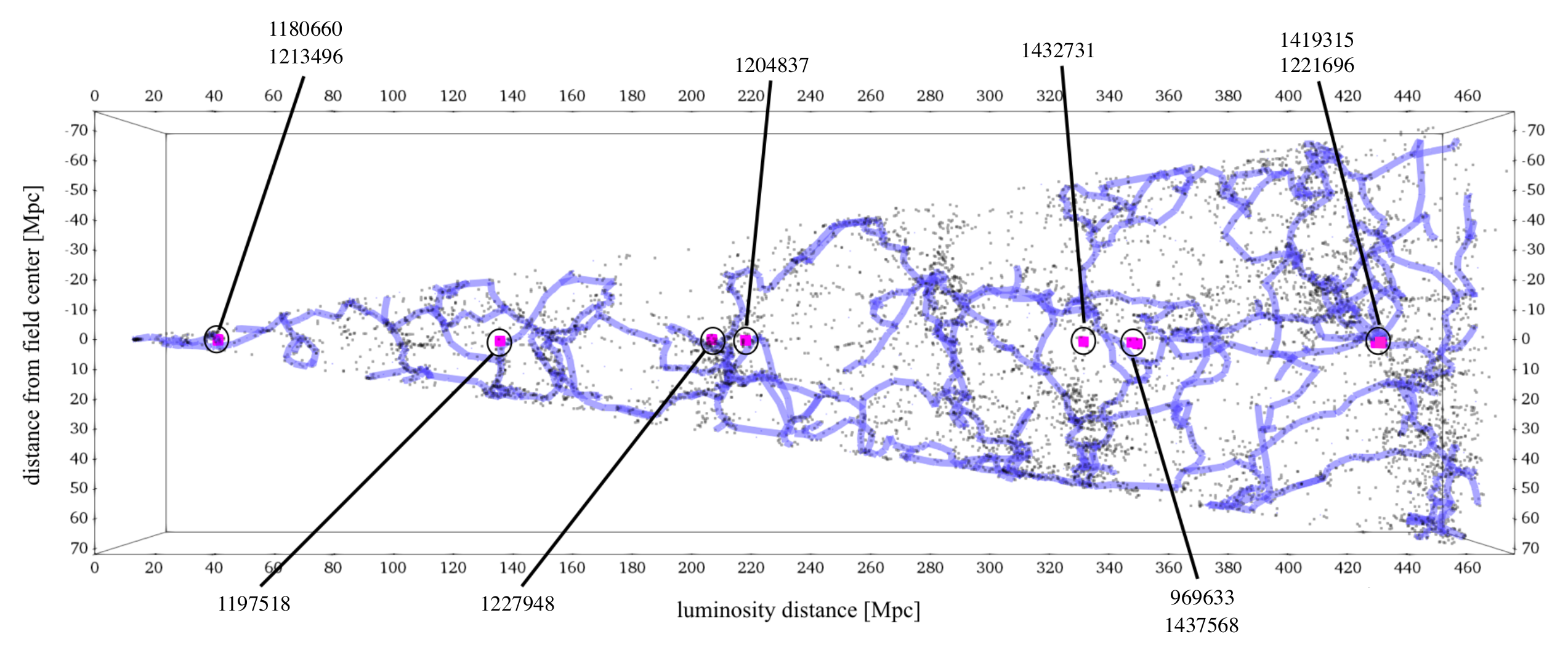}
\caption{Filamentary network of the cosmic web, based on the DisPeSE topological algorithm, overlaid on the distribution of galaxies in the redshift range 0 $<$ z $<$ 0.1. The black squares represent SDSS DR14 galaxies with known optical spectroscopic redshifts and the pink squares show the galaxies in the HI sample. At the edges of the figure, the filaments are mostly parallel to the edge of the sample. This is an artifact of the filament detection, due to galaxy density rapidly dropping at the survey edges. Note that the CHILES field of view is a slim cone with an extent of 5 Mpc at z = 0.1.}
\label{figure1}
\end{figure*}

\subsection{SALT Observation and Data Reduction}
The optical long-slit data is obtained (see Table 2 for observation properties) with the Robert Stobie Spectrograph (RSS) on the 11-meter SALT telescope between May 2016 and April 2017 under programs 2016-1-SCI-020, 2016-2-SCI-045, and 2017-1-SCI-047 (PI: J. Davis). The 2300 line mm$^{-1}$ volume phase holographic grating (R $\sim$ 4500, velocity resolution 68 km s$^{-1}$ for wavelength regions in this paper) with a 1.5" slit is used for galaxies up to z = 0.067, after which H$\alpha$ shifts off the CCD. The 1800 line mm$^{-1}$ volume phase holographic grating (R $\sim$ 4000, velocity resolution 71 km s$^{-1}$ for wavelength regions in this paper) with a 1.25" slit is used for the rest of the galaxies. The narrower slit for the latter observations is selected to aid in sky subtraction, as OH line complexes become increasingly dense after $\sim$ 7000 \AA. The RSS CCD has a pixel scale of 0.1267"/pix, and we employed 2$\times$2 binning for a pixel size of 0.2534". Seeing at the SALT telescope site for the observations varied between $\sim$ 1" and 1.5".

SALT is a fixed altitude telescope with its instrument payload located on a prime-focus tracker. Observations of most objects are limited to windows averaging 50 minutes twice per night. With this limitation in mind, exposure times are optimized to obtain acceptable signal-to-noise in the desired optical emission lines without spreading an observation over multiple nights. When possible, observations are fit into a single track to minimize sky-subtraction complications arising from changing flexure of the optical elements in the prime focus instrument payload.

When granted sufficient observing time, two slit position angles are selected for each galaxy; otherwise, one position angle (PA) is used. Primary position angles are selected to align with the major axis of the galaxy as identified in NED using r-band SDSS isophotal or K$_{s}$ values where available. However, in some cases, acquisition of galaxies within the slit required alignment with a bright star. In these cases, the slit is aligned as closely as possible to the major axis. The secondary slit position is selected to be $\sim$ 45$^{\circ}$ offset from the major axis or aligned with any potentially interesting optical features as seen in Hubble Space Telescope ACS images of the target galaxies.   

Reduction of the SALT data is carried out using the PySALT software, a package that implements standard pyraf procedures for SALT imaging and spectroscopic data \citep{Crawford2010}. We note that the spectra are not flux calibrated, as observations of a flux calibrator added prohibitive amounts of observing time and are deemed unnecessary for our kinematic analysis. The 2D spectra are sky-subtracted and wavelength-calibrated, with exposures combined when available, resulting in signal-to-noise ratios of $\sim$ 5-50 across the emission lines of interest. All galaxies but one exhibited H$\alpha$ $\lambda$6563 \AA, and seven galaxies exhibited one or more of the forbidden emission lines [N II] $\lambda$ 6583 \AA~ and the doublet [S II] $\lambda\lambda$ 6716, 6732. We note however that the sky subtraction for SALT at times leaves heavy residuals due to variable curvature in the skylines across the CCD, so often the [SII] doublet profile is damaged by intervening OH emission lines.

\section{Sample}
Our sample of galaxies is drawn from HI detections found in the CHILES pilot survey and narrowed down to ten galaxies within our H$\alpha$ observational limits on SALT. From the HI detections in the CHILES sample, galaxies with V-band magnitudes less than 20 and angular sizes greater than 10" are selected to ensure reasonable exposure times and sufficient spatial information for kinematic analysis. The redshift limit is set by the lowest resolution grating we are willing to use -- the PG1800 (R $\sim$ 4000, v $\sim$ 70 km s$^{-1}$) -- for which H$\alpha$ shifts off the CCD at z = 0.35, the presence of heavy CCD fringing effects beyond 8000 \AA, and the limiting magnitude of m$_{V}$ $<$ 20 for reasonable exposure time. We note that, because we are using H$\alpha$ and HI emission, we are necessarily biased towards star-forming, gas-rich objects. 

\begin{figure*}
\includegraphics[scale=0.58]{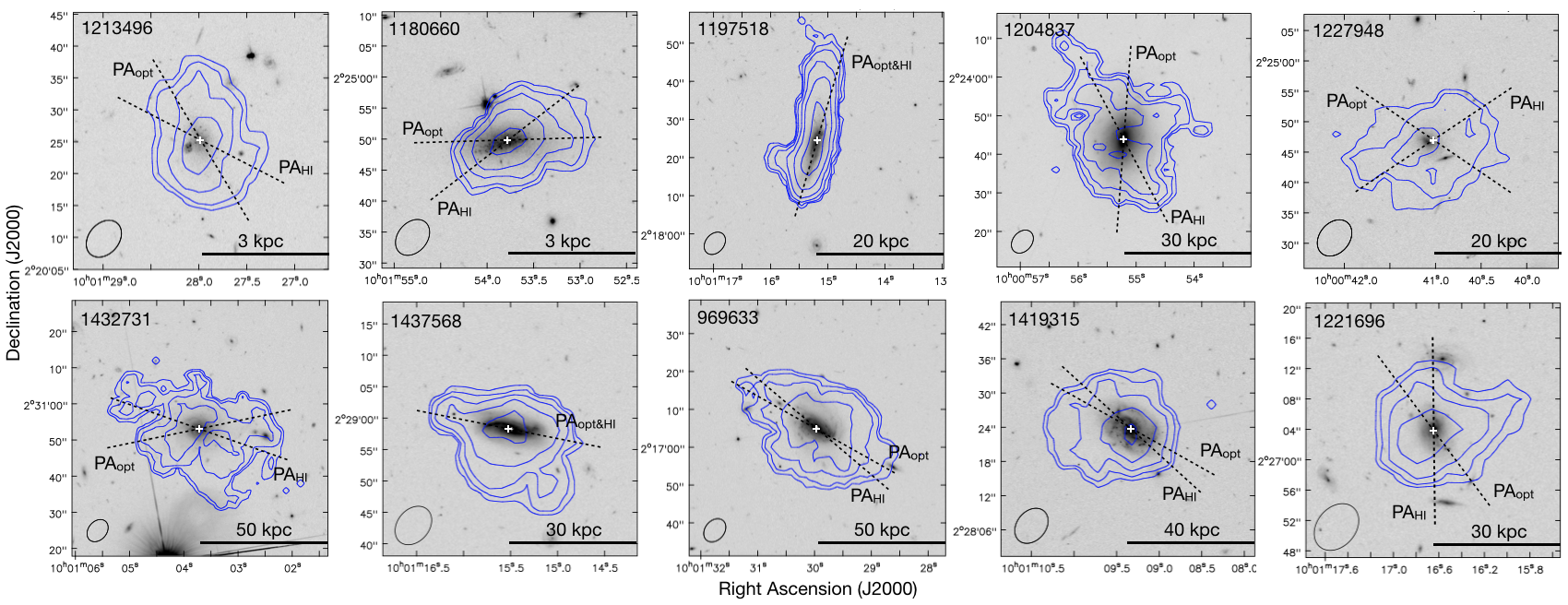}
\caption{Total HI intensity maps of the sample. The total integrated flux contours are 2, 4, 8, 16, and 32$\sigma$. The contour values in cm$^{-2}$ are listed in the Appendix (Figures A1 - A10) for each galaxy.}
\label{figure2}
\end{figure*}

\begin{figure*}
\includegraphics[scale=0.63]{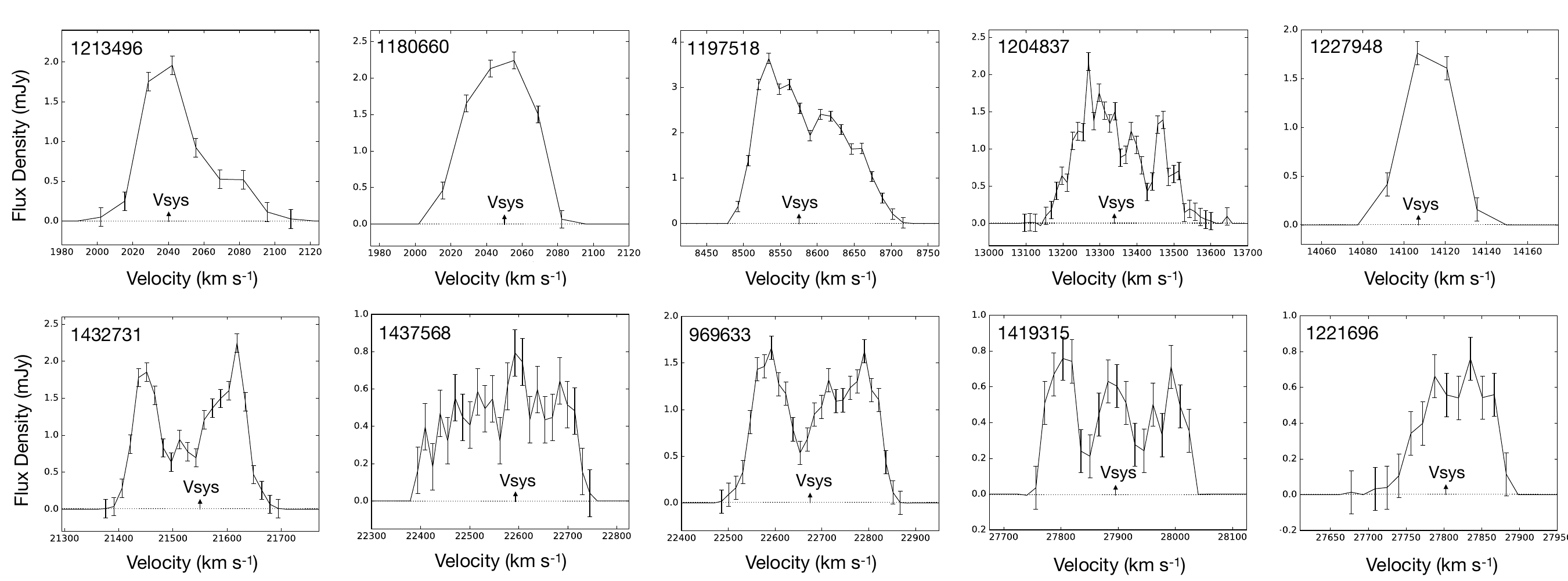}
\caption{Global HI profiles for the sample. The HI system velocity is indicated with an upward-pointing arrow on the profile.\;\;\;\;\;\;\;\;\;\;\;\;\;\;\;\;\;\;\;\;\;\;\;\;\;\;\;\;\;\;\;\;\;\;\;\;\;\;\;\;\;\;\;\;\;\;}
\label{figure3}
\end{figure*}

Our sample of ten galaxies is comprised of various morphological types. They are located in a range of different environments and redshifts as shown in Figure 1 and Table 3. The galaxy types include two dwarf irregulars (1213496 \& 1180660), three irregulars (1197518, 1227948, \& 1221696), three spirals (1204837, 1432731, \& 1419315) and two barred spirals (1437568 \& 969633). The redshift of the galaxies ranges from z = 0.0068 to 0.0930 with corresponding distances of 30 to 443 Mpc. The two closest galaxies (1213496 \& 1180660) may be gravitationally interacting and the two farthest galaxies (1419315, 1221696) may be gravitationally interacting as well. The galaxies have HI masses that range from $1.8\times10^{7}$ to $1.1\times10^{10}$ M$_{\odot}$, shown in Table 4. Half of the galaxies have HI extending nearly 2 - 3 times the optical radius, using the SDSS r-band isophotal major axis. One galaxy (1432731) has HI extending to nearly six times the optical radius. Most of the sample exhibits irregularities in the morphology and kinematics of their gas disks. Of the ten galaxies, six have the HI PA offset from the optical PA by 30$^{\circ}$ or greater, discussed further in Section 5.1. The HI properties are given in Tables 3 and 4. Total HI intensity maps of the entire sample are shown in Figure 2. Global HI profiles of the entire sample are shown in Figure 3.

\subsection{Stellar Properties}
Multi-wavelength coverage from COSMOS provides data from which we determine stellar properties. Stellar masses and SFR are estimated from SED fitting on UV through IR data using the G10/COSMOS v05 catalog \citep{Andrews2017} with the energy balanced SED fit program MAGPHYS \citep{daCunha2008} and are shown in Table 4. None of the galaxies in our sample have stellar masses above $3.0\times10^{10}$ M$_{\odot}$, the transition mass identified in observations below which galaxies are typically younger and in the assembly process \citep{Kauffmann2003}. The galaxies have stellar masses that range from $5.8\times10^{5}$ to $1.5\times10^{10}$ M$_{\odot}$. The uncertainty in the stellar mass is on the order of 17\%, except for the smallest masses. As \citet{daCunha2008} note, the smallest galaxies are not well fit with their method. Comparing results with different methods suggests that stellar masses for our two smallest galaxies are highly uncertain. The highest SFR in our sample is $\sim$2 M$_{\odot}$ yr$^{-1}$ with the majority falling below 1 M$_{\odot}$ yr$^{-1}$. The uncertainty in the SFR is on the order of 23\%.

\begin{table*}
\begin{flushleft}
\caption{Galaxy Properties of the Sample}
\begin{tabular}{ccccccccccc}
\hline
Galaxy& COSMOS& R.A.& Decl.& HI (Optical)& Dist.& V$_{sys}$& Type& Dist.Near& Dist.Fil& Fig.\\
ID& 08 ID& [J2000]& [J2000]& Redshift& [Mpc]& [km s$^{-1}$]& & [Mpc]& [Mpc]&\\
(1)& (2)& (3)& (4)& (5)& (6)& (7)& (8)& (9)& (10)& (11)\\
\hline
J100128.00+022025.4 &1213496 &150.3666 &2.3403 &0.0068 (0.0069) &30 &2041 &dIrr &0.173 &0.295 $\pm$ 0.089 &A1\\
J100153.77+022449.8 &1180660 &150.4741 &2.4139 &0.0068 (0.0068) &30 &2046 &dIrr &0.173 &0.319 $\pm$ 0.062 &A2\\
J100115.19+021824.4 &1197518 &150.3130 &2.3068 &0.0286 (0.0266) &130 &8575 &Irr &1.5 &3.7 $\pm$ 0.1 &A3\\
J100055.21+022343.8 &1204837 &150.2300 &2.3955 &0.0445 (0.0446) &205 &13335 &Sp &0.371 &3.4 $\pm$ 0.3 &A4\\
J100041.07+022446.7 &1227948 &150.1711 &2.4130 &0.0470 (0.0425) &217 &14113 &Irr &0.371 &3.6 $\pm$ 0.3 &A5\\
J100103.70+023053.1 &1432731 &150.2654 &2.5148 &0.0718 (0.0714) &337 &21554 &Sp &1.2 &9.6 $\pm$ 0.2 &A6\\
J100115.50+022858.5 &1437568 &150.3145 &2.4829 &0.0753 (0.0752) &354 &22596 &bSp &1.7 &7.1 $\pm$ 3.5 &A7\\
J100130.00+021705.0 &969633 &150.3748 &2.2848 &0.0756 (0.0750) &356 &22676 &bSp &1.7 &6.0 $\pm$ 3.4 &A8\\
J100109.33+022823.8 &1419315 &150.2889 &2.4732 &0.0930 (0.0927) &443 &27896 &Sp &0.265 &5.5 $\pm$ 0.5 &A9\\
J100116.64+022704.0 &1221696 &150.3194 &2.4511 &0.0927 (0.0927) &441 &27802 &Irr &0.265 &5.7 $\pm$ 0.4 &A10\\
\hline
\multicolumn{11}{p{17.1cm}}{\footnotesize{(1) Galaxy ID; (2) COSMOS 08 ID; (3) Units of right ascension are in degrees; (4) Units of declination are in degrees; (5) HI redshift is from CHILES and optical redshifts are from the G10/COSMOS v05 catalog; (6) Distance to target galaxy; (7) System velocity; (8) Galaxy morphological classification, done by eye; (9) Distance to nearest neighbor; (10) Distance to nearest filament; (11) Figure in the Appendix.}}\\
\end{tabular}
\label{table3}
\end{flushleft}
\end{table*}

\begin{table*}
\begin{flushleft}
\caption{Stellar and HI Properties of the Sample}
\begin{tabular}{ccccccccccc}
\hline
COSMOS& RMS& Beam& Abs. Mag.& NUV-r& SFR& M$_{*}$& M$_{HI}$& M$_{HI}$/M$_{*}$& D$_{Opt}$& D$_{HI}$\\ 08 ID& [$\mu Jy$ bm$^{-1}$]& [arcsec$^{2}$]& [W]& & [M$_{\odot}$ yr$^{-1}$]& $10^9$ [M$_{\odot}$]& $10^9$ [M$_{\odot}$]& & [kpc]& [kpc]\\
(1)& (2)& (3)& (4)& (5)& (6)& (7)& (8)& (9)& (10)& (11)\\
\hline
1213496 &76 &$6.42\times4.74$ &-14.0 &1.8 &0.002 &\ 0.0006 &0.018 $\pm$ 0.006 &30.7 &1.3 &1.4\\
1180660 &76 &$6.42\times4.74$ &-15.1 &1.4 &0.02 &\ 0.02 &0.029 $\pm$ 0.005	&1.3 &3.0 &2.3\\
1197518 &74 &$6.49\times4.83$ &- &1.8 &0.05 &\ 0.4 &1.7 $\pm$ 0.2 &4.2 &13 &22\\
1204837 &79 &$6.60\times4.93$ &-21.6 &4.3 &0.2 &13.3 &4.2 $\pm$ 1.4	&0.3 &27 &30\\
1227948 &79 &$6.60\times4.93$ &- &1.8 &0.03 &\ 0.2 &0.9 $\pm$ 0.3 &5.3 &7.3 &19\\
1432731 &86 &$6.77\times5.08$ &-19.7 &1.0 &0.6 &\ 8.2 &11.2 $\pm$ 3.1 &1.4 &10 &63\\
1437568 &86 &$6.77\times5.08$ &-20.8 &3.0 &1.9 &14.8 &5.2 $\pm$ 3.0	&0.4 &23 &39\\
969633 &86 &$6.77\times5.08$ &-21.2 &2.6	&1.6 &11.5 &9.6 $\pm$ 2.9 &0.8 &25 &47\\
1419315 &83 &$6.80\times5.10$ &-20.7 &2.4 &2.4 &\ 7.6 &6.3 $\pm$ 3.8	&0.8 &26 &34\\
1221696 &83 &$6.80\times5.10$ &-19.5 &1.3 &0.9 &\ 2.5 &3.2 $\pm$ 2.5	&1.3 &16 &12\\
\hline
\multicolumn{11}{p{16.4cm}}{\footnotesize{(1) COSMOS 08 ID; (2) Mean RMS of the image cube; (3) Synthesized beam FWHM of the image cube; (4) SDSS z-band absolute magnitude; (5) NUV-r, using GALEX NUV magnitude and SDSS DR7 r-band magnitude; (6) SFR; (7) Stellar mass; (8) HI mass, corrected for the primary beam; (9) Gas fraction; (10) Optical diameter along the SDSS r-band isophotal major axis; (11) HI diameter along the HI major axis, corrected for the beam width.}}\\
\end{tabular}
\label{table4}
\end{flushleft}
\end{table*}

\begin{table*}
\begin{flushleft}
\caption{Stellar, HI, and H$\alpha$ Properties of the Sample}
\begin{tabular}{cccccccccccc}
\hline
COSMOS& Incl.& W$_{int}$& W$_{20}$& W$_{pvd}$& W$_{H{\alpha}}$& PA$_{opt}$& PA$_{H{\alpha}}$& PA$_{HI}$& PA Offset& Spin$_{*}$& Spin$_{HI}$\\ 
08 ID& [deg]& [km s$^{-1}$]& [km s$^{-1}$]& [km s$^{-1}$]& [km s$^{-1}$]& [deg]& [deg]& [deg]& [deg]& Diff [deg]& Diff [deg]\\
(1)& (2)& (3)& (4)& (5)& (6)& (7)& (8)& (9)& (10)& (11)& (12)\\
\hline
1213496 &39 &120 &69 &55 &NA &30 &NA &242 &32 &0 &32\\
1180660 &65 &80 &63 &80 &80 &91 &84 &127 &36 &61 &83\\
1197518 &72 &244 &180 &190 &160 &165 &163 &163 &2 &3 &5\\
1204837 &43 &509 &323 &355 &135 &-5 &175 &207 &32 &42 &10\\
1227948 &40 &69 &42 &45 &40 &55 &54 &303 &68 &18 &86\\
1432731&14 &324 &222 &240 &240 &101 &101 &70 &31 &21 &10\\
1437568 &61 &353 &311 &350 &300 &78 &258 &258 &0 &54 &54\\
969633 &25 &395 &287 &350 &280 &60 &240 &229 &11 &72 &83\\
1419315 &49 &300 &248 &265 &230 &60 &240 &227 &13 &19 &6\\
1221696 &49 &215 &125 &150 &120 &36 &180 &180 &36 &5 &41\\
\hline
\multicolumn{12}{p{15.8cm}}{\footnotesize{(1) COSMOS 08 ID; (2) Inclination is calculated such that 0$^{\circ}$ is face-on; (3) HI line width over which the global HI profile is integrated; (4) HI line width measured at 20\% of the peak flux density; (5) HI line width measured at the maximum velocity of the rising and declining parts of the HI PV diagram; (6) H$\alpha$ line width measured at the maximum velocity of the rising and declining parts of the H$\alpha$ rotation curve; (7) Optical PA; (8) H$\alpha$ PA; (9) HI PA; (10) Difference between the optical and HI PAs; (11) Difference between the stellar spin angle and filament angle;  (12) Difference between the HI spin angle and filament angle.}} \\
\end{tabular}
\label{table5}
\end{flushleft}
\end{table*}

The galaxies in the sample are moderately inclined at 39$^{\circ}$ $<$ i $<$ 72$^{\circ}$ with two galaxies having more face-on inclinations of 25$^{\circ}$ and 14$^{\circ}$. The inclinations, given in Table 5, are calculated from SDSS DR7 \citep{Abazajian2009} r-band isophotal major and minor axes using the formula: 
\begin{equation} \label{eq1}
%\begin{split}
\sin{i} = \sqrt{ \frac{(1-(b/a)^2 )}{(1-q_{o}^2)} }
%\end{split}
\end{equation}
where a and b are the major and minor axes, and q$_{o}$ = 0.2 is the three-dimensional axis ratio following \citet{Huang2012}.

Star-forming galaxies follow a tight correlation between their stellar mass (M$_{*}$) and SFR \citep{Schiminovich2007}, with smaller galaxies having a lower SFR. This correlation has a bimodal distribution of blue (late-type) actively star-forming galaxies and of red (early-type) galaxies with little or no current star formation. The population of galaxies shifts from blue to red near a stellar mass transition of $3\times10^{10}$ M$_{\odot}$ \citep{Kauffmann2003}. In our sample of galaxies, SFR increases with increasing stellar mass (Table 4).

There is a similar bimodality in the mass dependence of the SFR per unit stellar mass. Figure 4 shows the relation of specific star formation rate (SSFR) versus stellar mass for our sample. The blue line shows the fit to the star-forming sequence for an SDSS spectroscopic sample of galaxies using GALEX ultraviolet luminosities to measure the SFR \citep{Schiminovich2007}. Except for one, our sample appears to lie along the blue sequence. There is no clear trend when examined as a function of redshift. The SSFR increases with decreasing galaxy mass, implying that lower mass galaxies form a higher fraction of their stellar mass in the present time.

\begin{figure}
\includegraphics[scale=1.2]{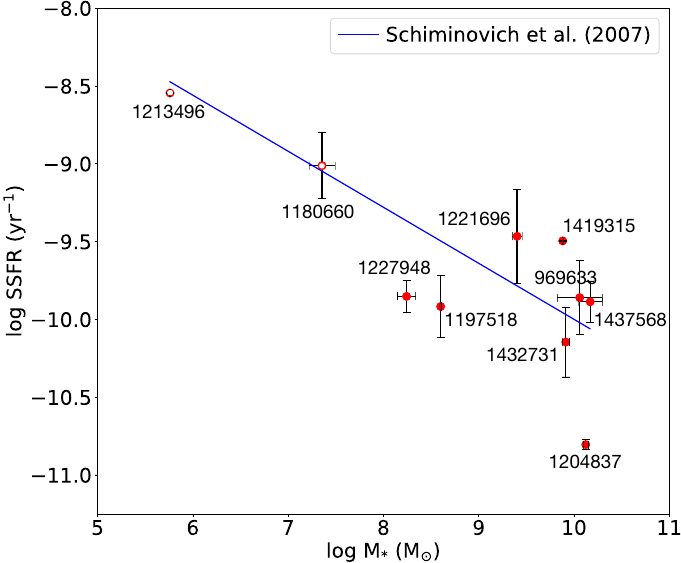}
\caption{Relation of the SSFR and stellar mass. The blue line shows the fit to the star-forming sequence for an SDSS spectroscopic sample of galaxies using GALEX ultraviolet luminosities to measure the SFR in \citet{Schiminovich2007}. Open symbols are small galaxies with uncertain stellar masses.}
\label{figure4}
\end{figure}

This bimodality is not absolute, with a green valley between the red and blue sequences. This valley consists of less active star-forming galaxies representing a combination of inactive disk galaxies and active bulge-dominated galaxies. The one outlier in our sample, 1204837, appears to lie in this green valley. In the HST image, this galaxy appears to be a bulge-dominated system, which has a high stellar mass, a lower SFR, and a NUV-r of 4.3.

\subsection{HI Data Analysis}
Sub-cubes with individual galaxies, described in Section 2.1, are used to produce total HI intensity maps (zeroth moment) and velocity fields (first moment) using SoFiA, the Source Finding Application \citep{Serra2015}. A noise scaling filter is applied along the velocity axis to normalize the cube by the local noise level per channel to account for variable noise characteristics throughout the cube. The S$\&$C algorithm \citep{Serra2012} is used to search for emission at multiple resolutions by smoothing the cube in three dimensions with specified kernels. The cube is smoothed at two resolutions in the sky using Gaussian kernels. The kernels are $4\times4$ and $6\times6$ pixels, which equal roughly 1.5, and 2 times the synthesized beam respectively. The cube is smoothed at multiple resolutions in velocity using boxcar kernels. The kernels vary from 3 to 11 times the channel width, with the combination depending on the width of the spectrum of the galaxy. At each resolution, a specified relative flux threshold (in multiples of the noise level) is applied, to extract and mark the significant pixels on each scale. The threshold varies from 3.5 to 4 for moment 0 maps. Higher threshold ranges of 4 to 6 are used for moment 1 maps as they are more sensitive to the noise. A final mask is produced through the union of the masks constructed at the various resolutions, with significant pixels merged into this final source mask that is then applied to the input cube.

We present figures for individual galaxies in the Appendix (Figures A1 - A10). The total HI intensity or moment 0 maps of each galaxy are overlaid as contours on Hubble Space Telescope (HST) Advanced Camera for Surveys (ACS) I-band (F814W) mosaic images (Figure 2, Figure 13, \& Figures A1 - A10) obtained from the COSMOS Archive and IRSA cutout service \citep{Koekemoer2007, Massey2010}. The optical center is marked with a white cross. The HI major axis is shown as a dotted line with the PA of the receding side. The optical major axis is also shown as a dotted line. The HI intensity-weighted velocity fields or moment 1 maps are similarly overlaid as contours on HST images. On the velocity field, the line passing through the cross represents the HI system velocity, V$_{HI}$. All nearby background galaxies with known redshifts in the HST images are confirmed to be distant enough to be ruled out of association or interaction with our targets. Only one target (1221696) has a neighboring galaxy's emission detected in the SALT spectra, but it is $>$ 30,000 km s$^{-1}$ in separation.  

The two-dimensional position-velocity (PV) slices are extracted along the HI major (${\phi}$) axis and minor (${\phi}$ + 90) axis. The optical center and the HI PA, shown in the upper-left corner, are used to make slices. The optical center of the slice and the HI system velocity are indicated with dashed lines. The global HI profiles for the sample (Figure 3 \& Figures A1 - A10) are produced by summing the flux in each channel using the source masks produced by SoFiA. Details on determining the HI system velocity and the PA for the HI major axis are found in Section 3.4. The contour levels of the HI emission in the PV diagrams, the HI column density in the total HI intensity maps, and the HI velocities of the HI intensity-weighted velocity fields are listed in the captions of Figures A1 - A10 in the Appendix.

\subsection{H$\alpha$ Data Analysis}
The ionized gas rotation curves are constructed using Python curve-fitting routines to fit single Gaussians to the emission line features in the 2D SALT spectra. A Gaussian is fit to each pixel row in the region of the emission line, sampling the spatial extent of the galaxy image on the CCD until the signal became too low to fit. From the optimized curve-fit, a continuum fit, amplitude, wavelength center, and FWHM is obtained. We use the wavelength centers to calculate a velocity. The spatial centers are assigned by finding the midpoint row of the emission profile. We note that by sampling the velocity centers at each pixel row, we are over-sampling with respect to the seeing (0.25" vs. $\sim$ 1"), but even when averaged, the rotation curve shape remains intact. Additionally, we note that while single Gaussians fit most of the emission line data well, two galaxies (1197518 and 1419315) exhibited profiles that are not well-fit by a single Gaussian. Details on the fitting of these two galaxies can be found in the sample descriptions in Section 3.6.

For seven of the galaxies in the sample, the optical SALT spectra exhibited one or both lines of the doublets [N II] $\lambda\lambda$ 6548,6583 \AA~ and [S II] $\lambda\lambda$ 6716, 6732 \AA. Though not necessary for the kinematic analysis, we use these emission lines to derive line ratios across the slit-region of each galaxy to search for any potentially interesting features in the ionized gas. The [N II] and [SII] lines are fit with the same method as described for H$\alpha$. The optimal continuum fit is then subtracted before calculating the area under the fit for H$\alpha$, [N II], and [SII]. The ratios [NII]$\lambda$6583/H$\alpha$ and [SII]$\lambda$ (6716 + 6732)/H$\alpha$ are then calculated row-by-row, sampling the ratio across the galaxy. One-sigma errors are derived from the Gaussian fits and propagated. Due to heavy residuals from the subtraction of night skylines, it is not always possible to fit one or both of the [SII] doublet lines. Given that the [SII] doublet lines are often comparable in intensity, for the galaxy in which we could not fit both lines, we make a very rough approximation of simply doubling the value derived from the fitted line. 

\begin{table}
\caption{Line Ratio Summary}
\begin{tabular}{ccccccc}
\hline
C08& NII/H$\alpha$& SII/H$\alpha$& NII/H$\alpha$& SII/H$\alpha$& OIII/H$\beta$\\
ID& Obs$^{a}$& Obs& SDSS$^{b}$& SDSS& SDSS\\
\hline
1197518 &0.07$\pm$0.02 &-- &-- &-- &--\\
1204837 &0.43$\pm$1.20 &-- &0.85 &0.95 &1.30\\
1432731 &0.28$\pm$0.03 &0.29$\pm$0.03 &0.31 &0.35 &0.30\\
1437568 &0.40$\pm$0.01 &0.33$\pm$0.03 &0.41 &0.37 &0.43\\
969633 &0.51$\pm$0.12 &0.34$\pm$0.12 &0.60 &0.26 &0.36 \\
1419315 &0.33$\pm$0.06 &-- &-- &-- & --\\
1221696 &0.15$\pm$0.03 &0.51$\pm$0.09 &-- &-- &--\\
\hline
\multicolumn{6}{p{8cm}}{\footnotesize{$^{a}$Median value of all fit points across target $^{b}$Values for 3'' diameter fiber centered on target.}}
\end{tabular}
\label{table1}
\end{table}

The median of the line ratios across each disk is compared to Sloan Digital Sky Survey (SDSS) values and found to be in decent agreement for all targets for which SDSS data is available, as shown in Table 6. All but three galaxies in our sample exhibited some or all of these forbidden lines, representing galaxies with SFRs between 0.05 and 2.4 M$_{\odot}$ yr$^{-1}$. While the median line ratio values for these galaxies agree with SDSS values, we find that there is some variation in the values across the disks. The [NII] emission in 1204837, the galaxy with the lowest SFR of the galaxies with detectable [NII] emission is too weak for spatially-resolved study, but for the other six galaxies, we are able to trace the line ratio across the inner disk (Figure 5). Only the innermost disk (out to radii of 3" to 6") is measured (see plots in Figures A1 - A10 for galaxy images). None of the galaxies for which there are SDSS O[III] data are beyond the star-forming region of a Baldwin-Phillips-Terlevich (BPT) diagram \citep{Baldwin1981}. As a final probe, we apply the simple metallicity scaling relationship developed by \citet{Dopita2016} across the disk (Figure 5). This relationship uses a ratio of [NII]/[SII] and [NII]/H$\alpha$ to obtain a 12 + log(O/H) metallicity value without the use of oxygen lines, which suffer from reddening effects or go unobserved without multiple spectrograph configurations. For the galaxies which contained the necessary nitrogen and sulphur lines, we find that, like the [NII]/H$\alpha$ ratio, the metallicity peaks in the center and decreases with increasing radius in the disk, as expected from typically negative radial metallicity gradients found in the disks of late type galaxies \citep{Marino2013,Belfiore2017}. The line ratio properties of each target are discussed in section 3.6.

\begin{figure*}
\includegraphics[scale=0.668]{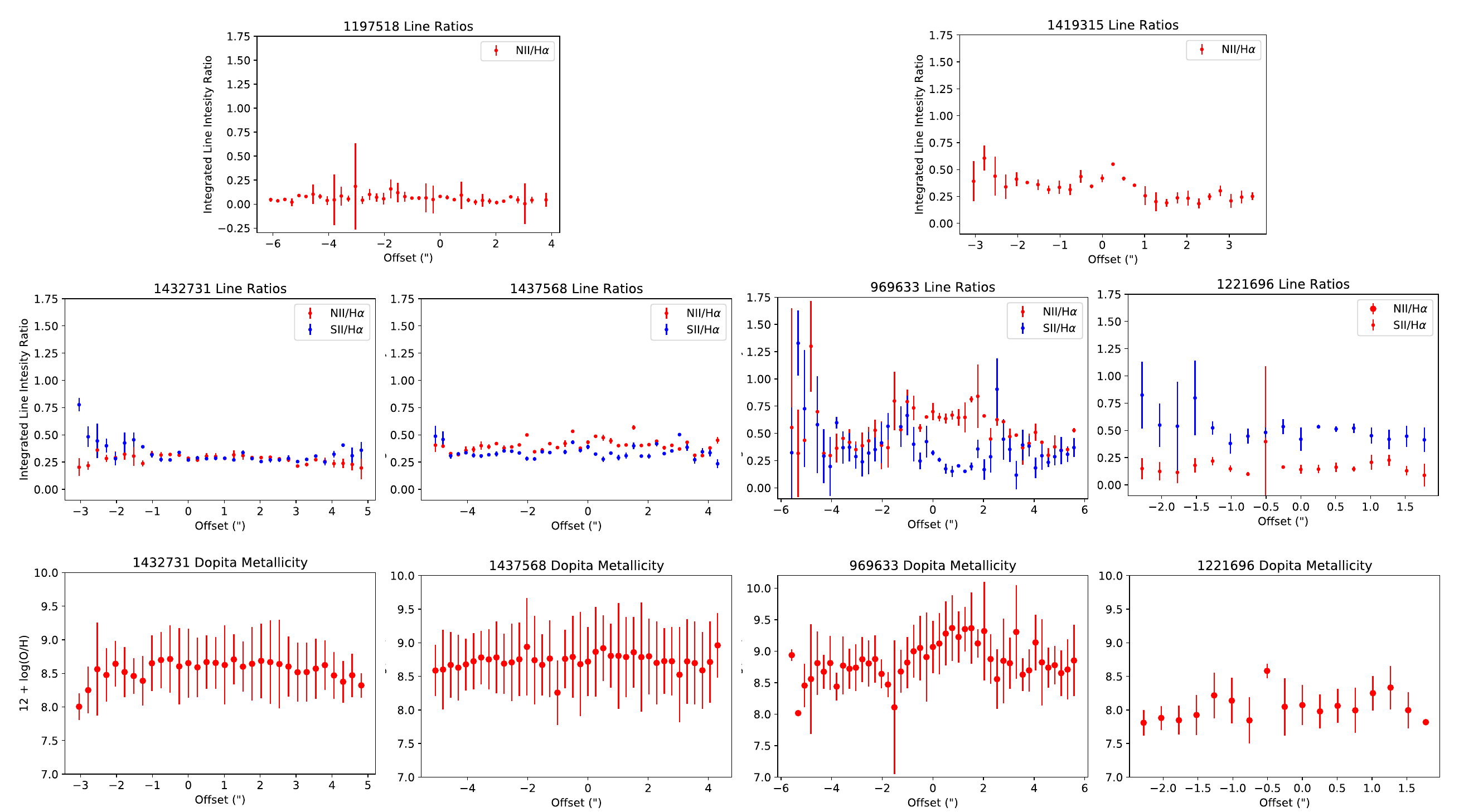}
\caption{Top and middle row: Line ratios of [NII]$\lambda$6583/H$\alpha$ (red) and [SII]$\lambda$ (6716 + 6732)/H$\alpha$ (blue) as a function of offset from the galaxy center for all galaxies in which these emission lines are measurable. Bottom row: Metallicity across the disk of the galaxy using the relationship from \citet{Dopita2016}.}
\label{figure5}
\end{figure*}

\subsection{HI and H$\alpha$ Properties}
HI properties for the galaxies are listed in Tables 3 and 4. The HI mass is calculated as: 
\begin{equation} \label{eq2}
%\begin{split}
M_{HI} = 49.8\:d{^2} \int s(\nu) d\nu \:[M_{\odot}]
%\end{split}
\end{equation}
where d is the luminosity distance in Mpc, s is the flux density in Jy, $\nu$ is the frequency in Hz, and flux is the integral of s d$\nu$ in $Jy \; Hz$. The integrated flux is determined as the sum of all flux density values within the source mask generated by SoFiA. This value is multiplied by the spectral channel width and divided by the number of pixels per beam to get flux in units of $Jy \; Hz$. The 5$\sigma$ HI mass sensitivity of our observations over 150 km s$^{-1}$ is $1.9\times10^{7}$ M$_{\odot}$ at z = 0.0068, and $3.5\times10^{9}$ M$_{\odot}$ at z = 0.0930. Our sample galaxies have HI masses ranging from $1.8\times10^{7}$ (velocity width 120 km s$^{-1}$) to $1.1\times10^{10}$ M$_{\odot}$, with six galaxies between $10^{9}-10^{10}$ M$_{\odot}$ and 1 galaxy above $10^{10} M_{\odot}$. The HI masses have been corrected for the primary beam. Throughout the HI cubes, we used the areas outside of the HI emission to measure the RMS. We estimate the HI mass to have an uncertainty on the order of 20\%. 

The column densities for individual galaxies are given in the captions of Figures 2 and A1 - A10 in the Appendix. The HI column density is calculated as:
\begin{equation} \label{eq3}
%\begin{split}
N_{HI} = \frac{2.34\times10^{20}}{\theta_{1}\theta_{2}} \:(1+z)^{4} \int s(\nu) d\nu \:[cm^{-2}]
%\end{split}
\end{equation}
where $ \theta_{1}$ and $\theta_{2}$ are the FWHM of the major and minor axes of the synthesized beam in arcsec, z is the redshift, s is the flux density in Jy, $\nu$ is frequency in Hz, and flux is the integral of s d$\nu$ in $Jy \; Hz$. We reach the theoretical noise in our image cubes and reach the predicted column density level of $3\times10^{19}$ cm$^{-2}$ (13 km s$^{-1}$ channel at z $\sim$ 0). 

Additional HI properties listed in Tables 3 and 4 are the system velocity, the line width of the galaxy, the radial extent of the HI, and the PA of the HI. V$_{sys}$ is taken as the velocity value at the optical center of the velocity field. W$_{int}$ is the line width over which the global HI profile is integrated and is taken from the channel range of the source mask used to generate moment 0 images in SoFiA. W$_{20}$ is the line width measured at 20\% of the peak flux density. W$_{pvd}$ is the line width measured at the maximum velocity of the rising and declining parts of the HI PV diagram. Similarly, W$_{H{\alpha}}$ is the line width measured at the maximum velocity of the rising and declining parts of the H$\alpha$ rotation curve. Errors of 27 km s$^{-1}$ in the HI line widths reflect uncertainties of one channel on either side. 

The radial extent of the HI diameter D$_{HI}$ is measured along the HI major axis of the PV diagram at a limiting column density of $1.25\times10^{20}$ cm$^{-2}$ (1 M$_{\odot}$ pc$^{-2}$). D$_{HI}$ is corrected for beam smearing effects using a Gaussian approximation \citep{Wang2016}:
\begin{equation} \label{eq7}
D_{HI}  = \sqrt{ (D_{HIo}^2) - (B^2) }
\end{equation}
where D$_{HI}$ and D$_{HIo}$ are the corrected and uncorrected HI diameters, and B is the synthesized beam along the major axis. Errors of 11'' in the HI radial extent reflect uncertainties of one beam-width on either side. 

PA$_{HI}$ is calculated using SoFiA to determine the flux-weighted centroid of the emission in each channel of the image cube and then fitting a straight line to the set of centroids. Errors of 10$^{\circ}$ in the HI PA reflect an uncertainty estimate of 5$^{\circ}$ on either side. The PA is compared by eye to the HI kinematic major axis of the velocity field. The PA is adjusted to match the HI kinematic major axis of the velocity field in 1432731 and 1221696. PA$_{H{\alpha}}$ is taken at the optical major axis or close to it, as described in Section 2.2. 

\subsection{Identifying Galaxy Environments}
To quantify the environments of galaxies, we look at their location in the cosmic web and the distance to the nearest neighbor. \citet{Luber2019} has developed the use of DisPerSE for the CHILES volume using a catalog of 11500 spectroscopic redshifts from the G10/COSMOS v04 catalog \citep{Davies2015}. They show that for this small volume sensible results are obtained that are consistent with larger surveys. Here we use the same method, but we use redshifts from SDSS DR14 rather than from G10/COSMOS v04 since too few redshifts are available at this low redshift in G10/COSMOS v04. We search SDSS DR14 for all galaxies with spectroscopic redshifts with coordinates 148$^{\circ}$ $<$ R.A. $<$ 153$^{\circ}$, and 0$^{\circ}$ $<$ Dec. $<$ 5$^{\circ}$ which corresponds to a thickness of 70 Mpc at the higher end of the redshift range and 15 Mpc at the lower end of the redshift range. We choose an area that is sufficiently wider than the actual CHILES field, to properly reconstruct the large-scale structure in the CHILES field. 

We run DisPerSE over this galaxy catalog with a mirror boundary condition and a significance level of four. See \citet{Luber2019} Section 4 for further explanation of the method and these experimentally determined parameters. DisPerSE identifies various types of structures in a density field of points by characterizing topological features which correlate to indexed manifolds which then correspond to various components of the cosmic web (voids, walls, filaments and clusters). We extract the filament spines, with the tracers of this filamentary structure referred to as critical points. We calculate the distance of a galaxy to its nearest filament spine by finding the distance of that galaxy from the nearest critical point. The error in distance is bound by half the distance between the critical point used and the next closest one. Figure 1 shows the three-dimensional slice of our field of view collapsed into a two-dimensional projection. The position of galaxies in relation to the filamentary structure is also shown in Figure 1 and discussed in Section 5. Note that contrary to some other work where two-dimensional distances are used \citep{Kleiner2017, Laigle2018, Luber2019}, we use three-dimensional distances in our analysis.

The distance to the nearest neighbor is calculated by assuming all velocities are purely due to expansion, converting RA, DEC, and redshift to physical units, and finding which galaxy is the minimum distance in physical units. It is important to note that this is done with the spectroscopic catalog as photometric redshifts are not accurate enough for low z. 

\subsection{Notes on Individual Galaxies}
\textbf{1213496 \& 1180660} ---
The two nearest galaxies in our sample are dwarf irregulars (Figures A1 \& A2). They are close spatially with a separation of around 70 kpc, as shown in Figure 6a, and have a velocity difference of 5 km s$^{-1}$. Both 1213496 and 1180660 lie within $\sim$300 kpc of filamentary structure. These galaxies have HI masses on the order of 10$^{7}$ M$_{\odot}$ and have the highest SSFR of our sample. Both galaxies have HI offset from the optical major axis as well as asymmetric HI morphology. In each galaxy, the optical PA follows the outer HI intensity contours while the HI PA follows the velocity field of the inner disk. For 1213496, the northern part of the velocity field in the outer disk seems to shift more towards polar. For 1180660, the velocity field looks even more disturbed with the contours to the west nearly perpendicular to the HI major axis, more like a polar ring of counter-rotating gas. Given their proximity, the disturbances in the outer disks may be an indication that these two galaxies are gravitationally interacting.

\begin{figure}
\includegraphics[scale=1.1]{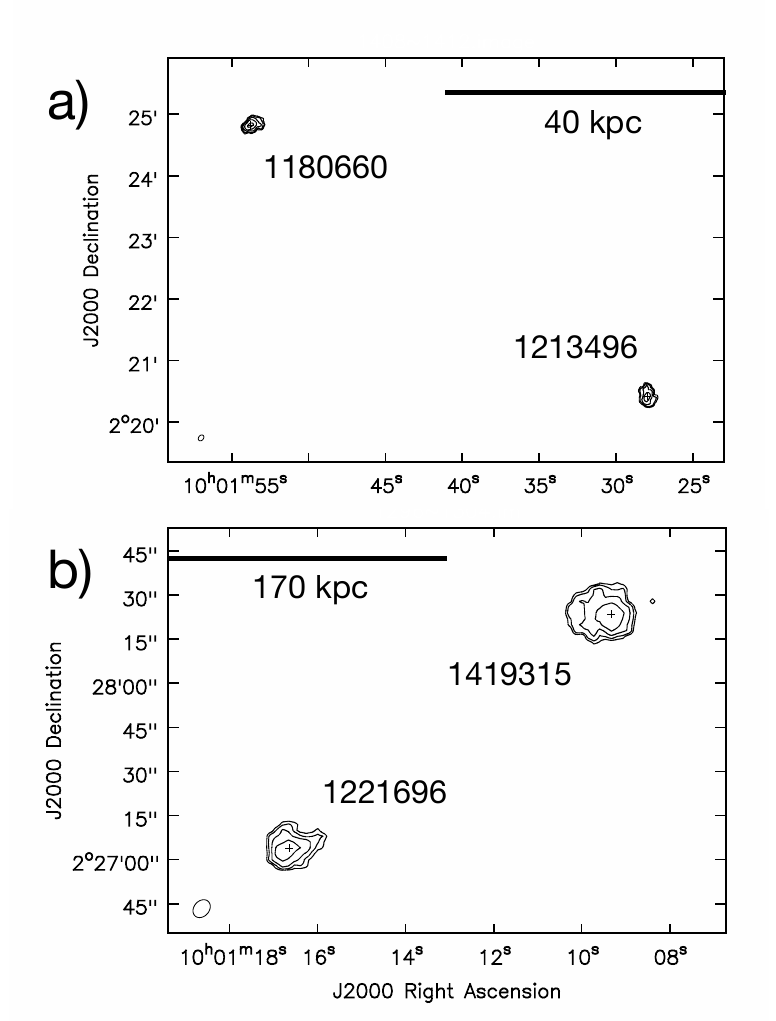}
\caption{Galaxies in close proximity. Figure 6a shows total HI intensity maps for the two nearest galaxies in our sample. They are close spatially with a separation of around 70 kpc and have a velocity difference of 5 km s$^{-1}$. For both galaxies, the HI velocity field of the outer disk shifts more towards being perpendicular to the HI major axis, indicative a polar ring. Their disturbed HI morphology and kinematics along with their proximity may indicate they are gravitationally interacting (Figures A1 \& A2). Figure 6b shows total HI intensity maps for the two farthest galaxies in our sample. They are close spatially with a separation of around 170 kpc and have a velocity difference of 94 km s$^{-1}$. 1221696 has a tail that extends in the direction of 1419315. This HI feature along with their proximity may indicate they are gravitationally interacting (Figures A9 \& A10).}
\label{figure6}
\end{figure}

Ionized gas is not detected in 1213496 and is limited in extent in 1180660, consistent with their low star formation rates. When viewing 1180660's ionized gas rotation overlaid on its HI PV diagram, we see that the ionized gas is well-aligned along with the system velocity for PA 84$^{\circ}$, and fits well within the HI contours for PA 137$^{\circ}$, though with a slightly steeper slope. 

\textbf{1419315 \& 1221696} ---
The two farthest galaxies in our sample are close spatially with a separation of around 170 kpc, and have a velocity difference of 94 km s$^{-1}$. 1221696 (Figure A10 \& Figure 6b) has a tail that extends in the direction of 1419315 (Figure A9). This feature along with their proximity may indicate that they are gravitationally interacting. Both 1419315 and 1221696 are located $\sim$5.6 Mpc from filamentary structure. These galaxies have HI and stellar mass on the order of 10$^{9}$ M$_{\odot}$ and have SFRs that are some of the highest in our sample. Both have asymmetric HI morphology.

The ionized gas rotation curves in both of these galaxies mostly agree with the HI PV diagram contours. However, in 1419315, the ionized gas rotation seems to flatten/decline at a lower velocity than the HI towards the receding edge of the galaxy. The HI and H$\alpha$ are decoupled at large radii, which is a unique phenomenon in this sample. Galaxy 1419315 did not have measurable [SII], but did demonstrate enhanced [NII]/H$\alpha$ ratio at the center, and asymmetric values (high on the approaching side and low on the receding side). This is consistent with a negative radial metallicity gradient. In galaxy 1221696, we again see little variation in the line ratios across the measured region of the disk. The [SII]/H$\alpha$ ratio is, however, higher at all points across the measured area of the disk. Given the limited region probed, it is difficult to draw any conclusions about these line ratio values. 

\textbf{1204837 \& 1227948} ---
1204837 (Figure A4) and 1227948 (Figure A5) are two galaxies that are within 371 kpc of their nearest neighbor. Both galaxies have extended HI with asymmetric morphology and are located $\sim$3.5 Mpc from filamentary structure. Both have HI offset from the optical major axis, with irregular galaxy 1227948 having a PA offset of 68$^{\circ}$.

The ionized gas rotation fits well with the HI PV diagram of 1204837, with both neutral and ionized gas missing from the center of the galaxy. 1204837 has the most dominant bulge in our sample, which may account for the lack of neutral and ionized gas in the center of the PV diagrams. 1204837 has very weak [NII] 6583 \AA~ emission. In 1227948 the ionized gas lays along with the system velocity for both position angles, agreeing well with the HI contours (albeit with some scatter since the galaxy is irregular and one of the smallest in the sample).

\textbf{1432731} ---
1432731 (Figure A6) is a spiral galaxy and is the most face-on galaxy of our sample, inclined at 14$^{\circ}$. It is at a distance of 1.2 Mpc from its nearest neighbor. This galaxy has the largest HI mass in the sample at $1.1\times10^{10}$ M$_{\odot}$ and resides 9.6 Mpc from a filament spine. 1432731 has an HI diameter that is 71 kpc across, extending over six times past the optical diameter. The HI distribution and morphology are very asymmetric and the PV diagram shows lower velocity low-level emission on the receding side which could indicate counter-rotation.

The ionized gas overall consistently traces the HI contours. There is a region of ionized gas emission on the receding edge, likely corresponding to a bright HII region in one of the faint extended spiral arms visible in the HST image of the galaxy. In 1432731, both [NII]/H$\alpha$ and [SII]/H$\alpha$ are similar values in the central region of the disk, but [NII] emission diminishes while [SII] emission increases at increasing radii. The decreased [NII] emission is consistent with decreasing metallicity. Though [SII]/H$\alpha$ may also trace metallicity, it does so much less reliably. When the [SII]/H$\alpha$ ratio is unusually high ($>$0.4), it can be attributed to supernova remnants (SNR) or diffuse ionized gas \citep{Zhang2017}. Our line widths do not seem wide enough for SNRs, so the enhanced [SII]/H$\alpha$ at the edges of the measurable region may therefore be attributable to diffuse ionized gas. Given that the "flared" [SII]/H$\alpha$ occurs around the point where the slit traverses inter-arm regions rather than spiral arms on both sides of the galaxy, this seems plausible. 

\textbf{1197518} ---
1197518 (Figure A3) is an irregular galaxy and one of the more isolated galaxies in our sample, at a distance of 1.5 Mpc from its nearest neighbor and 3.7 Mpc from a filament spine. It is the brightest HI detection so far in the full survey and has extended HI with very asymmetric morphology. The HI mass is several times its stellar mass and the SFR is low at 0.05 M$_{\odot}$ yr$^{-1}$.

The ionized gas kinematics for this galaxy are the most irregular in the sample. Weak [NII] 6583 \AA~ is present along with H$\alpha$. Though it fits within the HI contours, the rotation curve exhibits a central kink. In the outer regions of the galaxy (-10" to -7.5" and +2.5" to +10"), the emission line profiles are well fit by a single Gaussian. The central region is highly asymmetric, suggesting at least two components, with the dominant component shifting from red to blue as one moves from bottom to top of the galaxy spatially.

The steep rise and fall of the ionized gas in the central region may be indicative of a strong bar, though the galaxy seems to be viewed almost edge-on, making this conclusion difficult to confirm. \citet{Sofue1999} identified that, while barred galaxies exhibit similar general properties compared to unbarred galaxies, they have larger velocity amplitude variation in the innermost disk. This large velocity variation arises from the barred potential, and simulations of PV diagrams for edge-on barred galaxies show many tens of km s$^{-1}$ fluctuations, superposed on the usual flat rotation curve \citep{Athanassoula1999}.

Considering this is a highly inclined galaxy, the PV diagram can also be interpreted as to have a steeply rising rotation curve at the approaching side, and a more slowly rising rotation curve at the receding side, not inconsistent with the H$\alpha$. The optical morphology of the outer disk in the HST image of the galaxy seems suggestive of a warp. The HI contours have a mild curvature, but the ionized gas is more dramatically perturbed. With a nearest neighbor 1.5 Mpc away, a merging event seems unlikely but cannot be conclusively ruled out and lends a possible explanation for the disturbed kinematics of the ionized gas.

\textbf{1437568 \& 969633} ---
Two barred spirals are the most isolated galaxies in our sample, at a distance of 1.7 Mpc from their nearest neighbor. Both 1437568 (Figure A7) and 969633 (Figure A8) are located several Mpc from filamentary structure, at 7.1 Mpc and 6.0 Mpc respectively. These galaxies are large in stellar mass and have higher SFRs. Both galaxies have extended HI and 1437568 has asymmetric HI morphology.

The ionized gas emission in 969633 extends to 8" along PA = 60$^{\circ}$, corresponding to a distance of 19.9 kpc, or $\sim$ 80\% of the stellar radius (R$_{*}$ = 24.44 kpc). The profile of this galaxy differs from the rest of the sample due to the presence of a very broad, bright central region. The outer disk is fit by relatively narrow (FWHM $<$ 1 \AA~) Gaussians, but the central region has characteristic widths of 3 \AA, or 120 km s$^{-1}$. The galaxy is not, however, an AGN. Line ratios measured using H$\alpha$, [NII], and [SII] from the SALT spectrum, as well as SDSS spectrum values for [OIII]$\lambda$5007 / H$\beta$, place this galaxy firmly within the star-forming region of a BPT diagram \citep{Baldwin1981}. We thus attribute the broad line widths in the central region to the kinematic influence of the bar (see Section 3.3). The ionized gas matches the HI spatially and kinematically. The ionized gas rotation can be traced through the central region, where no HI is detected for the chosen position angle. The ionized gas in 1437568 is mostly unremarkable, following the HI contours in slope and velocity center. 

In galaxy 1437568, the line ratios remain relatively flat across the measured portion of the disk, albeit with more scatter than other galaxies in the sample. Though it is a barred spiral, we do not see a significant enhancement of [NII]/H$\alpha$ (see 969633 notes). The most interesting of the sample in terms of line ratios is 969633, with enhanced [NII]/H$\alpha$ and suppressed [SII]/H$\alpha$ in the central region. Given that this galaxy has the broadest line widths in the center in H$\alpha$, it would seem there is intense activity in the innermost part of the disk, though it does not qualify as an AGN. The enhanced [NII]/H$\alpha$ likely arises from the bar in 969633. \citet{Florido2015} find that barred galaxies exhibit an enhanced N/O ratio, and thus higher [NII]/H$\alpha$. They find no similar effect on [SII]/H$\alpha$, in agreement with the [SII]/H$\alpha$ values remaining mostly flat across the region of enhanced [NII]/H$\alpha$. 

\section{Scaling Relations}
We analyze ancillary stellar data along with the observed HI and H$\alpha$ properties of our sample and find our data follow known galaxy scaling relations. Our sample follows expected trends for star formation rates, HI gas fraction and the HI size-mass relation as well as the baryonic Tully-Fisher relation for the HI and H$\alpha$.

\subsection{HI Gas Fraction}
The relation between M$_{*}$ and M$_{HI}$ for our sample is shown in Figure 7. The diagonal grey solid line indicates equal amounts of stellar and HI mass. The average gas fraction (M$_{HI}$/M$_{*}$) of our sample is 1.8 (Table 4). This is excluding 1213496, which has a gas fraction of 31 and our highest SSFR. We compare our sample with the galaxy population detected by ALFALFA, the largest wide-field blind HI survey at low redshift \citep{Haynes2011}. The orange line shows the median values from the spectroscopic ALFALFA-SDSS galaxy sample in \citet{Maddox2015}. Our sample shows M$_{HI}$  increasing as a function of M$_{*}$ and agrees quite well with the median values of the ALFALFA sample. Galaxies with stellar masses below 10$^{9}$ M$_{\odot}$ have more HI than stars, and galaxies with stellar masses above 10$^{10}$ M$_{\odot}$ have less HI than stars. The galaxies in our sample with stellar masses below 10$^{9.5}$ M$_{\odot}$ are all classified as irregular.

\begin{figure}
\includegraphics[scale=1.2]{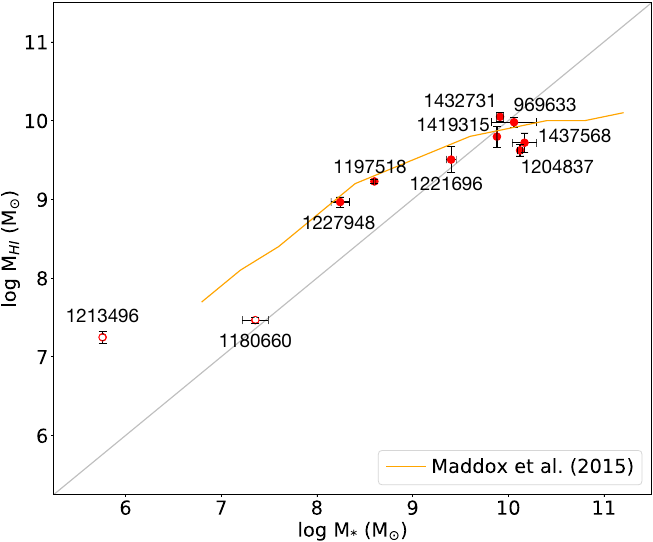}
\caption{Relation of the HI mass and stellar mass. The diagonal grey solid line indicates equal amounts of stellar and HI mass. The orange line shows the median values from the spectroscopic ALFALFA-SDSS galaxy sample in \citet{Maddox2015}. Open symbols are small galaxies with uncertain stellar masses.}
\label{figure7}
\end{figure}

\begin{figure}
\includegraphics[scale=1.1]{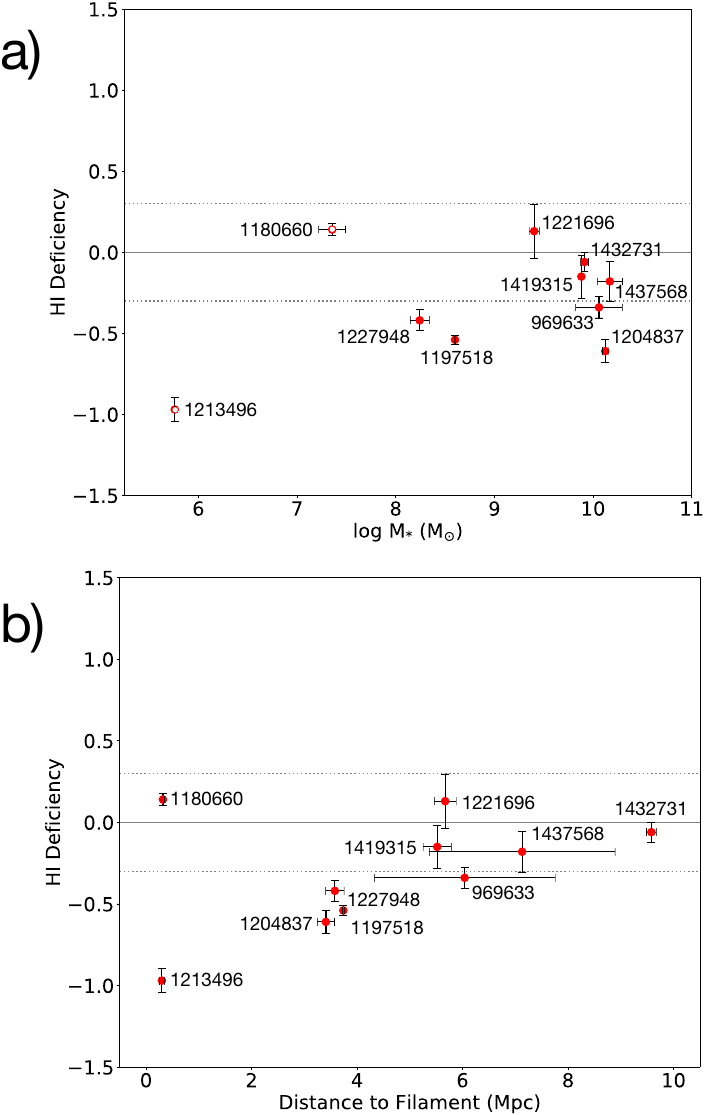}
\caption{HI deficiency using scaling relations from \citet{Catinella2012}. A solid grey line is drawn for deficiency = 0  and a dotted grey line for deficiency = +0.3 and -0.3. Figure 8a shows the HI deficiency in relation to the log of stellar mass. Figure 8b shows the HI deficiency in relation to the distance from filamentary structure based on the DisPeSE topological algorithm. Open symbols are small galaxies with uncertain stellar masses.}
\label{figure8}
\end{figure}

\subsection{HI Deficiency}
The HI content of galaxies can be characterized by scaling relations between the HI content and other intrinsic properties of galaxies. These HI scaling relations can be useful in identifying galaxies that have either more HI than expected or less HI than expected, as galaxies like these may have been affected by recent processes including the removal or accretion of gas. One way to characterize the HI gas content of galaxies is with HI deficiency. HI deficiency is the logarithmic difference between the observed HI mass and the expected HI mass, 
\begin{equation} \label{eq4}
%\begin{split}
{\rm HI_{def} = log\:M_{HIexp} - log\:M_{HIobs}}
%\end{split}
\end{equation}
where M$_{HIexp}$ is the expected HI mass calculated from scaling relations, and M$_{HIobs}$ is the measured HI mass taken from observations \citep{Haynes1984}. Generally, a galaxy is considered to have a normal HI gas content if its HI deficiency is between -0.3 and 0.3 \citep{Denes2014}, with an HI excess if $<$ -0.3 and HI deficiency if $>$ 0.3.

Studies use HI selected samples combined with SDSS optical properties to derive scaling relations. With the GASS survey, \citet{Catinella2012} found that a good predictor of the HI content is the linear combination of stellar surface density and NUV-r color. We examine the HI deficiency of our sample adapting results using the HI to stellar mass fraction function shown in Figure 8 from \citep{Catinella2012}$\colon$
\begin{equation} \label{eq5}
%\begin{split}
{\rm log\:M_{HI}/M_{*} = a\:log\:\mu_{*} + b\:(NUV-r) + c}
%\end{split}
\end{equation}
with a = -0.285, b = -0.366, c = 2.872. This function is the relation between HI mass fraction and a linear combination of stellar mass surface density and NUV-r color with the relation obtained using the subset galaxies with NUV-r $\leq$ 4.5 mag. The stellar mass surface density is calculated from the formula 
\begin{equation} \label{eq6}
%\begin{split}
{\rm \mu_{*} =  \frac{ M_{*} }{ 2\:\pi\:R_{50z}^{2} }\:\:[M_{\odot}\:kpc^{-2}] } 
%\end{split}
\end{equation}
where R$_{50z}$ is the radius containing 50\% of the Petrosian flux in z-band from SDSS DR7, and NUV-r is the GALEX \citep{Martin2005} NUV magnitude minus the SDSS DR7 r-band magnitude.

The results for HI deficiency are shown in Figure 8. A solid grey line is drawn for deficiency = 0 and a dotted grey line for deficiency = +0.3 and -0.3. Figures 8a and 8b show the HI deficiency in relation to stellar mass and distance to filament, with no obvious trends. In \citet{Catinella2012}, the results are for a sample of galaxies with stellar mass greater than $10^{10}$ M$_{\odot}$. Since only three galaxies (1204837, 1437568, 969633) are above stellar mass $10^{10}$ M$_{\odot}$, this method is not well calibrated for our lower mass sample. With that said, none of the galaxies are HI deficient and half of the sample has HI excess. This is of course not surprising for a small HI selected sample.

\begin{figure}
\includegraphics[scale=1.125]{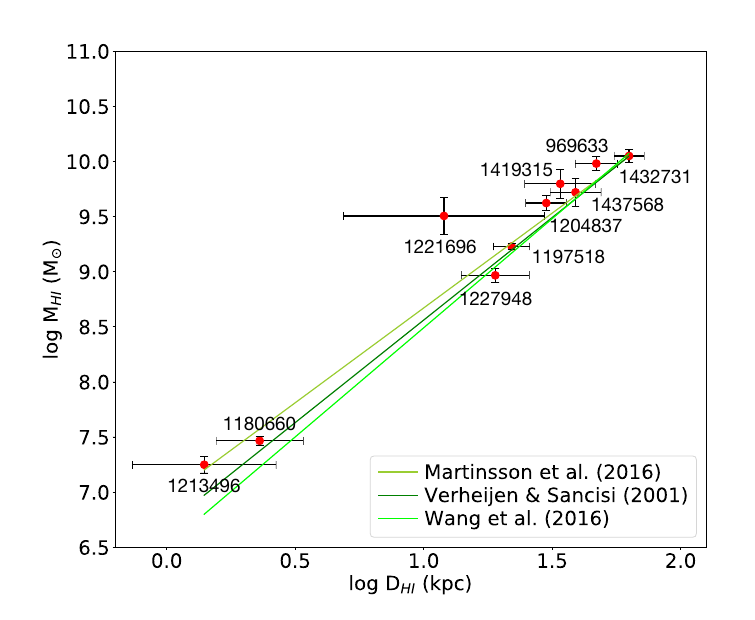}
\caption{Relation of the HI mass and HI disk diameter. To the radial extent of the HI diameter D$_{HI}$ is measured along the HI major axis of the PV diagrams at a limiting column density of $1.25\times10^{20}$ cm$^{-2}$ (1 M$_{\odot}$ pc$^{-2}$) and corrected for beam smearing. The lime green line is the correlation found by \citet{Wang2016}, the dark green line represents the correlation found by \citet{VerheijenSancisi2001} and the olive green line is the correlation found by \citet{Martinsson2016}. Note that these papers use azimuthally averaged surface density to measure D$_{HI}$ while we use a one dimensional PV diagram.}
\label{figure9}
\end{figure}

\subsection{HI Size-Mass Relation}
HI often extends beyond the optical disk, where it can trace events of removal and accretion of gas. Observationally, these events can be revealed by the asymmetries of the morphology and kinematics of HI \citep{Sancisi2008}. The majority of galaxies in our sample show extended HI. There is a tight correlation between the HI mass and the beam-corrected diameter of the HI disks, as shown in Figure 9. The radial extent of our HI diameters D$_{HI}$ is measured along the HI major axis of the PV diagram at a limiting column density of $1.25\times10^{20}$ cm$^{-2}$ (1 M$_{\odot}$ pc$^{-2}$). D$_{HI}$ is corrected for beam smearing effects. The correlation of HI mass and diameter in the plot shows that as disks become more massive, they grow in size and implies a nearly constant HI surface density regardless of size.

\citet{VerheijenSancisi2001} show, in their sample of 43 spirals, at least half have lopsided HI distribution and/or kinematics. Moreover, \citet{Swaters2002} examine 73 late-type dwarf galaxies and find that lopsidedness is as common among dwarf galaxies as it is in spiral galaxies. Furthermore, \citet{Jutte2013} conduct a statistical investigation of 76 HI disks finding at least 50 percent of galaxies have lopsided disks and that generally morphological and kinematic irregularities are correlated. Most of the galaxies in our sample have irregularities in the morphologies and kinematics of their gas disks and our results do not appear to be that unusual. 

\subsection{HI and H$\alpha$ Line Widths}
With our sample of galaxies, we study the structural relation of HI and H$\alpha$ disks with rotation curves of H$\alpha$ in the inner regions and with HI in the outer regions. The H$\alpha$ traces the kinematics at a spatial resolution of 1" and a velocity resolution of 70 km s$^{-1}$. The HI traces the kinematics at a spatial resolution of 5" and a velocity resolution of $\sim$ 13 km s$^{-1}$.

A comparison of HI and H$\alpha$ at low redshift is useful for studies at high redshift where only optical lines are observable. Studies at higher redshift provide information for scaling relations used in mass modeling of galaxies and in studying the evolution of the Tully-Fisher relation. We compare the line width of the HI versus the line width of the H$\alpha$ for our sample, as shown in Figure 10a. The H$\alpha$ PA is taken at the optical major axis or close to it, as described in Section 2.2. The HI PA is taken at the kinematic major axis, which differs from the optical major axis in more than half of the sample and may be a reason for the slight difference in widths between the HI and H$\alpha$. Except 1197518 and 1204837, the line width of the H$\alpha$ is of the same order as the line width of the HI. This indicates that the flat part of the rotation curve is reached for HI and H$\alpha$, even though the rotation curve may not completely flatten for all our galaxies in H$\alpha$. Uncertainty on the H$\alpha$ measurements has been visualized as a shaded red region on the PV diagrams (Figures A1 - A10), which represents the SALT velocity resolution of $\sim$ 70 km/s for the gratings used. This uncertainty is almost certainly an overestimate, as the emission-line profiles can be fitted with much higher accuracy. However, the errors on the Gaussian velocity centers are < 5 km/s for all points but the extreme edges, and thus nearly invisible when overlaid on the PV diagram.

\begin{figure}
\includegraphics[scale=1.1]{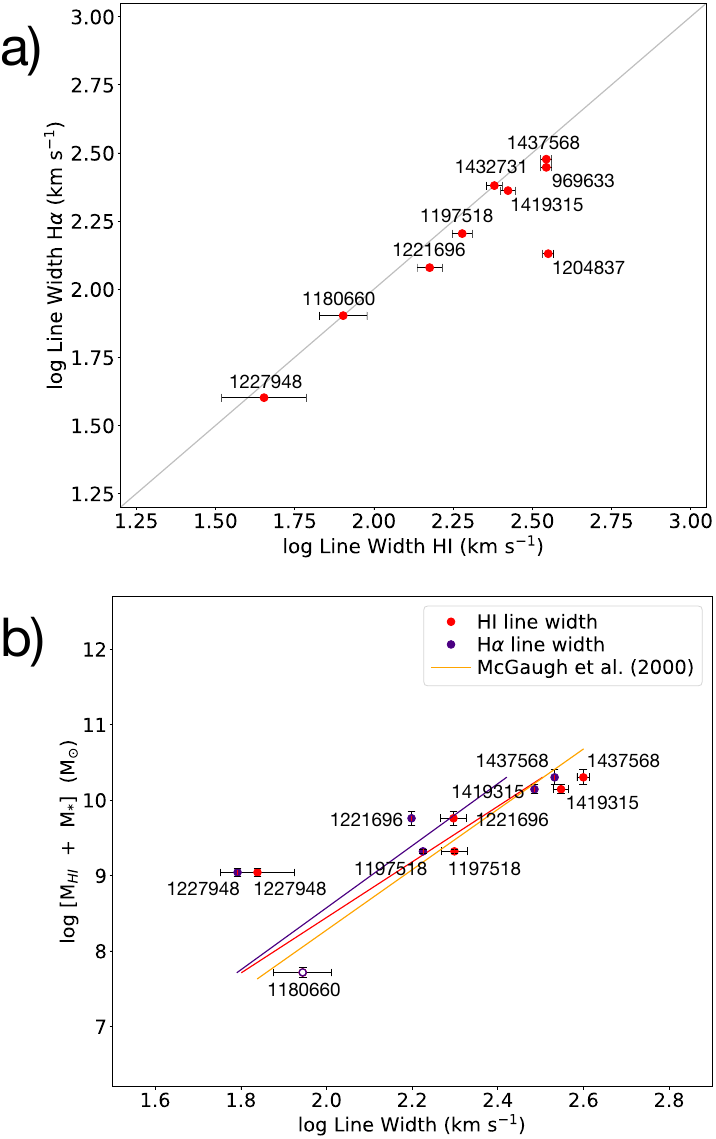}
\caption{Line widths of the neutral and ionized gas. Figure 10a shows the relation of the width of HI and the width of H$\alpha$. The diagonal grey solid line indicates equal widths of HI and H$\alpha$. Figure 10b shows the Baryonic Tully-Fisher Relation with total mass (M$_{HI}$ + M$_{\star}$) vs the width of HI and H$\alpha$. The BTFR from \citet{McGaugh2000} is shown in orange. Open symbols are small galaxies with uncertain stellar masses.}
\label{figure10}
\end{figure}

\citet{Pisano2001} compare line widths for HI and H$\alpha$ and find that for larger galaxies the line widths are similar but for smaller galaxies, there is a larger spread and the H$\alpha$ line width is usually smaller. We find that the HI and H$\alpha$ line widths are generally the same, with the exception of one galaxy. For galaxy 1204837, the H$\alpha$ PA of the slit is not aligned with the major axis and does not trace the full velocity width.

\subsection{HI and H$\alpha$ Baryonic Tully-Fisher Relation}
Despite the diverse formation histories of individual galaxies, local disk galaxies exhibit a tight relationship between their rotation velocity and their luminosity or mass, namely the Tully Fisher (TF) relation \citep{Tully1977}. Combining the stellar mass with observed gas mass results in a Baryonic Tully-Fisher Relation (BTFR) that is linear over many decades in mass \citep{Verheijen2001}. 

Figure 10b shows BTFR for our sample, with the total mass (M$_{HI}$ + M$_{\star}$) versus the line width of the HI and H$\alpha$. Line widths used in our Tully Fisher comparisons are corrected for inclination, with 1432731 (i=14$^{\circ}$) and 969633 (i=25$^{\circ}$) excluded from the relation. \citet{Lelli2019} study the BTFR at z = 0 using 153 galaxies from the SPARC sample with HI and H$\alpha$ rotation curves. They find the tightest BTFR is given by using the mean velocity along the flat part of the rotation curve with a best-fit slope of 3.85 $\pm$ 0.09. The slope of the line using the HI widths from our sample is 3.68 $\pm$ 0.66 and the slope of the line using the H$\alpha$ widths is 4.11 $\pm$ 0.83. This is excluding 1204837 as this galaxy is an outlier in Figure 10a, which causes a significant difference in the slopes for HI and H$\alpha$. Note that the slope should be interpreted with caution since there is a large scatter around this fit. The BTFR from \citet{McGaugh2000} is shown in orange for reference.

\section{Results}
We use ancillary stellar data and observed HI data to examine the role local environment plays in our sample by comparing galaxy properties with respect to the location of the nearest neighbor. In addition, we examine the role large-scale environment plays by comparing galaxy properties with respect to location and orientation of the nearest cosmic web filament.

\subsection{HI Morphology and Kinematics}
An indication in our data that neighbors may have an impact on the morphology is shown in Figure 11. Six out of ten galaxies have the PA offset between the HI kinematical major axis and the optical photometric major axis by 30$^{\circ}$ or greater. For our sample, the more offset the PA the closer the galaxy is to its nearest neighbor. Another indication that neighbors may impact morphology and kinematics is the two dwarf irregular galaxies (1213496, 1180660) shown in Figures 6a, A1, and A2. They have a separation in \text{the} sky of around 70 kpc and a difference in velocity of 5 km s$^{-1}$. For both galaxies, the HI velocity field of the outer disk shifts more towards being perpendicular to the HI major axis, indicative of a polar ring \citep{Stanonik2009}.

\begin{figure}
\includegraphics[scale=1.15]{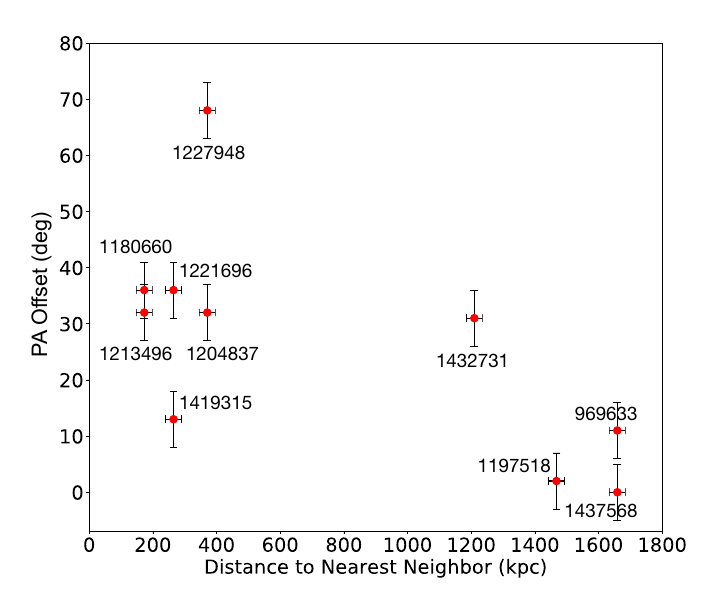}
\caption{Relation of the PA offset and distance to nearest neighbor. PA offset is the difference between the HI kinematical major axis and the optical photometric major axis.}
\label{figure11}
\end{figure}

There are several galaxies in the sample with more prominent irregularities in their HI morphology and/or kinematics. For example, 1432731 has the most disturbed HI distribution, 1437568 has the most asymmetric HI morphology, 1180660 has the most shifted HI kinematics compared to the optical disk, and 1204837 has the most noticeable HI PA misalignment compared to the optical PA plus two HI tails. Yet, the stellar disks of these galaxies appear to be undisturbed. There is no clear trend in HI irregularities with type, SFR, gas fraction, mass and environment of these galaxies as they all span the entire range of the sample. With the optical images showing undisturbed stellar disks, the HI properties depict a more complex scenario, one that may include the accretion of gas from the surrounding environment.

\subsection{Distance to Cosmic Web Filaments}
Both theory and observations suggest that large-scale structure impacts galaxy evolution in addition to known trends in local density. For example, observations show that at the same local density, redder, passive and more massive galaxies are found closer to their filaments \citep{Chen2017, Malavasi2017, Laigle2018, Kraljic2018}. Studies on HI content have mixed results. \citet{CroneOdekon2018}, using the ALFALFA catalog, find that HI deficiency decreases as the distance from the filament increases and that most gas-rich galaxies are in small tendrils within voids. In contrast, \citet{Kleiner2017}, using stacked spectra from the HIPASS catalog, find that for galaxies with larger stellar mass log M$_{*}$/M$_{\odot}$ $>$ 11, the HI fraction in the sample 0.7 Mpc from filaments is higher than the HI fraction in galaxies far away from filaments. The study finds no difference in HI fraction between galaxies at 0.7 and 5 Mpc from filaments at smaller stellar mass log M$_{*}$/M$_{\odot}$ $<$ 11.

Our sample of galaxies lies within a range of three-dimensional distances from the cosmic web. The two dwarf galaxies are around 300 kpc from a filament spine and the rest of the sample is from 3.4 to 9.6 Mpc from a filament spine. Since our low-mass sample is HI-biased, it can be expected that our galaxies may be more isolated \citep{Kreckel2012}. Figure 12a shows the HI mass increases with increasing distance from filaments. 1432731 is the most gas-rich galaxy in our sample with an HI mass of $1.1\times10^{10}$ M$_{\odot}$. It has an HI diameter that is 71 kpc across and resides at the furthest distance in our sample at 9.6 Mpc from a filament spine. Figure 12b shows the SSFR decreases with increasing distance from filaments. 1432731 has the lowest SSFR with respect to distance from filaments, with the exception of green valley galaxy 1204837. Since we have such a small sample, any of the relations shown should be considered only a hint of a trend at most. Lastly, the gas fraction in relation to distance from filamentary structure is shown in Figure 12c. Except for 1213496, 1227948 and 1197518, the gas fraction varies between 0.3 and 1.4 as a function of distance from filaments. Figures 12d, 12e, and 12f plot the same stellar and HI properties with respect to distances to the nearest neighbor and show no obvious trends.

\begin{figure*}
\includegraphics[scale=1.07]{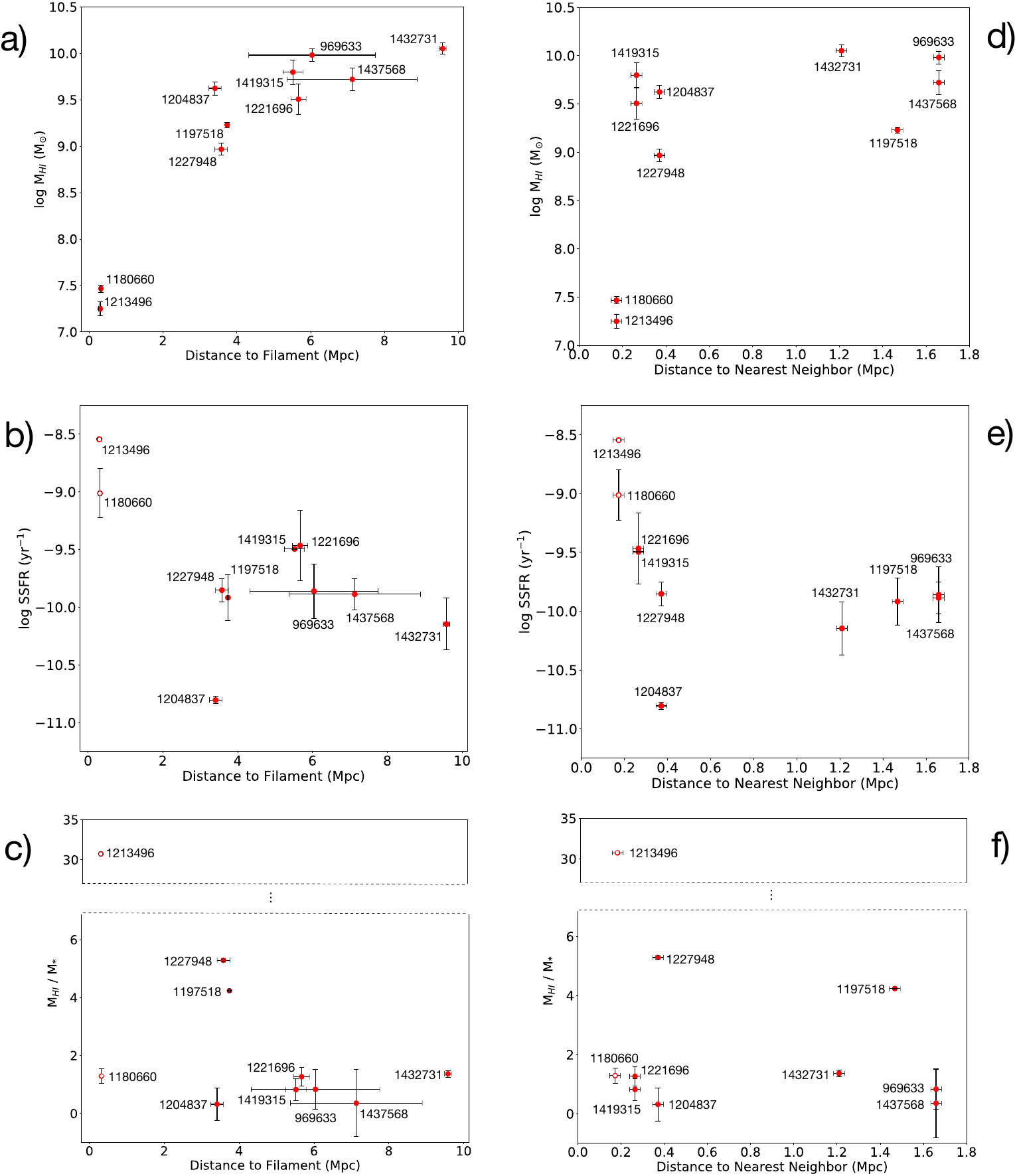}
\caption{Relation of the galaxy properties and galaxy location. Figure 12a shows the relation of HI mass and distance from the filament. Figure 12b shows the relation of SSFR and distance from the filament. Figure 12c shows the relation of gas fraction and distance from the filament. Figures 12d, 12e, 12f show the stellar and the HI properties of our sample with respect to the nearest neighbor. Figure 12d shows the relation of HI mass and distance from the nearest neighbor. Figure 12e shows the relation of gas fraction and distance from the nearest neighbor. Figure 12f shows the relation of SSFR and distance from the nearest neighbor. Open symbols are small galaxies with uncertain stellar masses.}
\label{figure12}
\end{figure*}

\begin{figure*}
\includegraphics[scale=0.85]{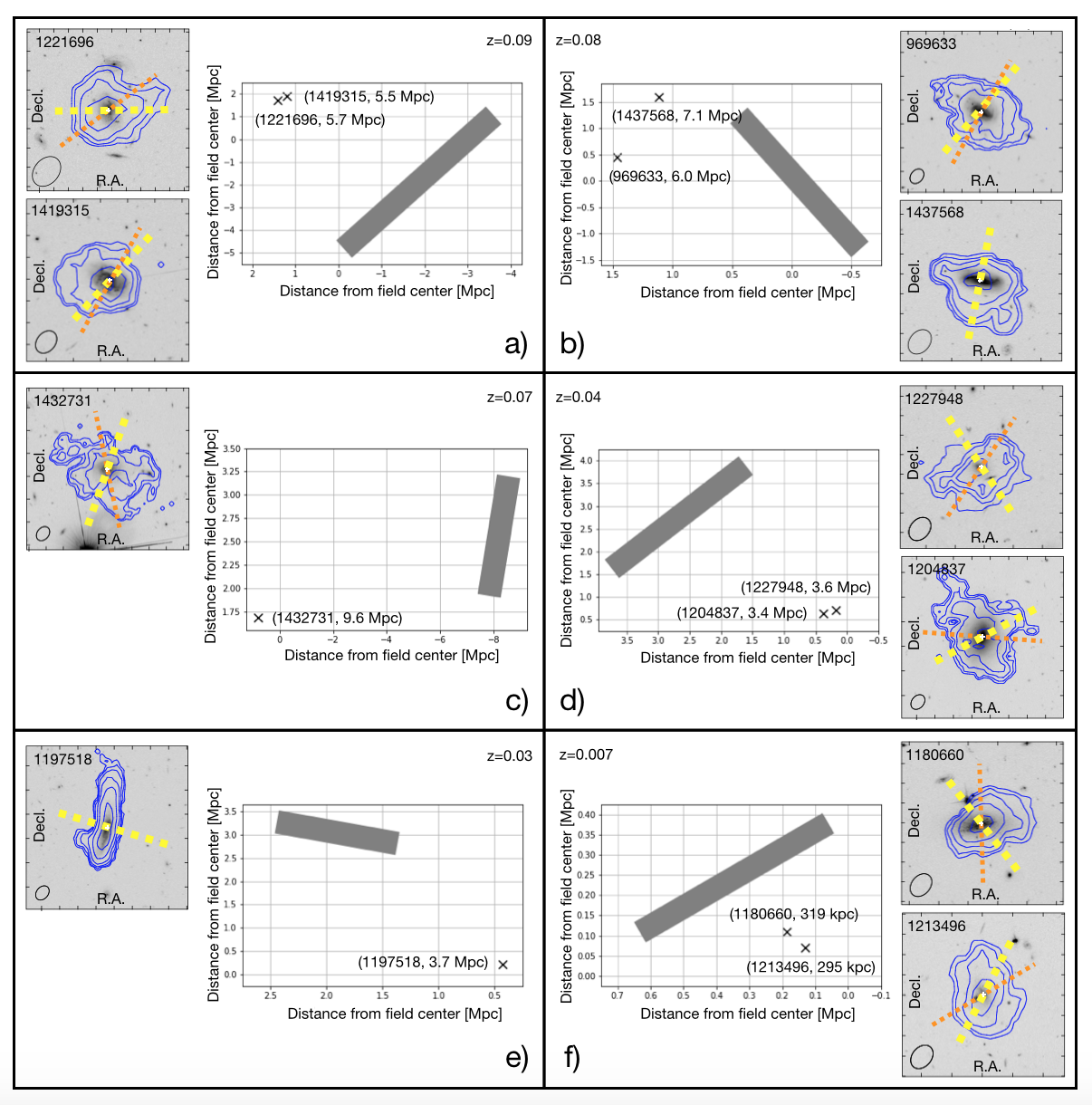}
\caption{Relation of the galaxy spin vector and its nearest filament. In each panel (a-f), the grey line highlights the filament that is closest to each galaxy (shown with an x). The closest filament is calculated in three dimensions and is found by selecting the three closest critical points. Once the closest filament is identified, it is projected in two dimensions onto the sky as shown on the plots in each panel. The plot axes are distance from the field center in Mpc (at the distance of the filament) and are comparable to right ascension and declination. The galaxy name and distance (in three dimensions) to the filament are shown next to each x. The redshift range is shown in the upper corner of each panel. Total HI intensity maps for the HI are shown along with the spin vectors for the HI gas (yellow-dashed line) and the spin vectors for the stellar disk (orange-dashed line) The axes are right ascension and declination.}
\label{figure13}
\end{figure*}

\subsection{Orientation to Cosmic Web Filaments}
Angular momentum is a key physical ingredient in galaxy formation and plays a crucial role in determining the history of a galaxy. Angular momentum has been a subject of classical investigations by \citet{Hoyle1949} and \citet{Peebles1969}. The work of \citet{Doroshkevich1970} and \citet{White1984} led to a standard theory for the origin of angular momentum in the framework of hierarchical cosmological structure formation known as tidal torque theory. Galaxy dark matter haloes form in over-densities of the cosmic web by accreting material and smaller galaxies via filaments \citep{Bond1996}. As large scale structure collapses, galaxies acquire angular momentum and this process imprints alignments between the spin of galaxies and surrounding filaments. Low-mass galaxies assemble by accreting onto filaments, generating spins that align with filaments. High-mass galaxies assemble by merging along filaments, generating spins perpendicular to filaments. Extension of the work on tidal torque theory \citep{Porciani2002} to non-linear environments \citep{Codis2015} predicts this spin flip transition of haloes as galaxies migrate along filaments and accumulate mass. This is also seen in increasingly detailed simulations \citep{Kraljic2019} predicting the spin alignment of galaxies and halos with respect to filaments, with spin flips occurring at a halo mass of $5\times10^{11}$ h$^{-1}$ to $5\times10^{12}$ h$^{-1}$ M$_{\odot}$.

Dark matter halo spin alignment has received much attention in the past, while studies of galaxy spin alignment have emerged only more recently. The dark matter halo spin flip transition is seen in both dark matter only simulations \citep{Wang2018, GaneshaiahVeena2018} and most hydrodynamical simulations \citep{Codis2018, GaneshaiahVeena2019}. The case of galaxy spin alignment is less clear. The galaxy spin flip transition is seen in hydrodynamical simulations \citep{Codis2018, Kraljic2019}. In contrast, \citet{GaneshaiahVeena2019} find a lack of detection of the galaxy spin flip transition and relate this to the filament thickness. The methods used in hydrodynamic simulations as well as the methods used to quantify the cosmic web may play a role in spin studies. Since \citet{Kraljic2019} use DisPerSE to identify the cosmic web in the SIMBA simulation \citep{Dave2019}, their results are directly comparable to our observations.

Observations of alignments between large-scale structure and the spin of galaxies show mixed results. \citet{Krolewski2019} find no clear evidence for alignment between galaxy spins and filament directions from kinematics using integral-field spectroscopy for 2700 galaxies from the MaNGA survey along with the Cosmic Web Reconstruction algorithm \citep{Chen2016} to identify filaments. However, hints of the spin flip for galaxies have been identified in SDSS using shape as a proxy for galaxy spin \citep{Tempel2013, Tempe&Libeskind2013, Pahwa2016, Chen2019}. More recently, \citet{Welker2019} detect mass dependent galactic spin alignments from kinematics using integral-field spectroscopy for 1278 galaxies from the SAMI survey, along with DisPerSE to characterize the cosmic web filaments.

\begin{figure}
\includegraphics[scale=0.31]{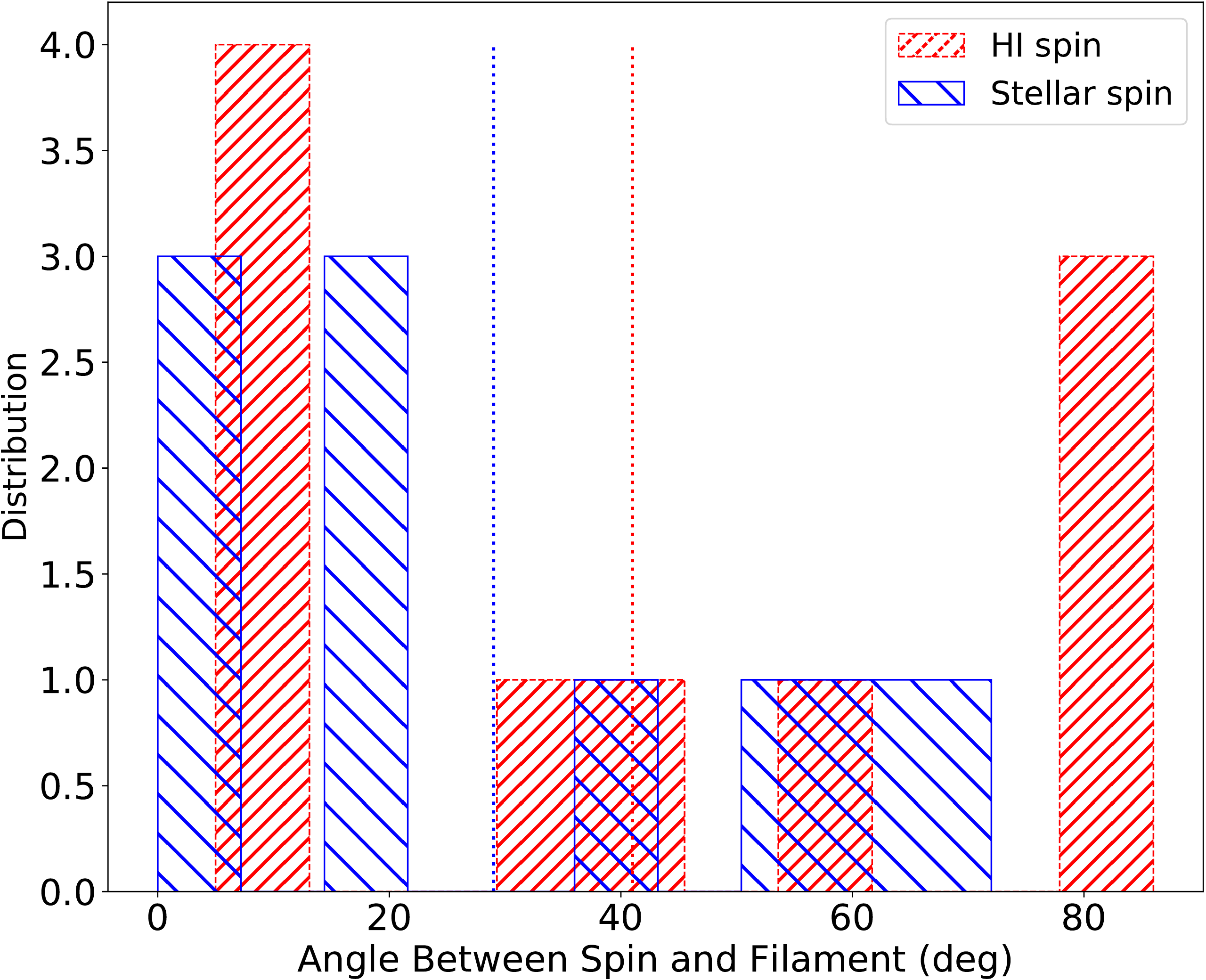}
\caption{Angle between the galaxy spin vector and its nearest filament. The red shows the spin determined from the HI gas and the blue shows the spin determined from the stellar disk. The dotted lines are the average value for HI and stellar spin.}
\label{figure14}
\end{figure}

We examine the spin vector alignments of our sample of galaxies with cosmic web filaments using the kinematic major axis of the HI gas and the photometric major axis of the stellar disk. The location of our sample of galaxies with respect to nearest filamentary spine based on the DisPerSE topological algorithm is shown in Figure 13. We measure the galaxy angular momentum vector or spin vector, from the PA of the major axis. The plane-of-sky projection of the spin vector is perpendicular to the PA of the major axis. The alignment between spin angle and filament angle is determined by taking the difference between the two apparent angles, both projected onto the sky.

Figure 14 shows a histogram of the angle between the galaxy spin and filament. The red shows the spins determined from the HI gas and the blue shows the spins determined from the stellar disk. We follow the convention that if the angle is smaller than 45$^{\circ}$, the alignment is referred to as parallel and conversely, if the angle is larger than 45$^{\circ}$, the alignment is referred to as perpendicular. There is spread in the alignment angles for both the stellar and HI components with a majority less than 45$^{\circ}$. The average spin angle for the HI component is 41$^{\circ}$ and the average spin angle for the stellar component is 29$^{\circ}$.

A transition between the aligned and perpendicular orientations of galaxy spins is found in some simulations, including the recent work by \citet{Kraljic2019}. They find a transition in orientation at a stellar mass of $\sim10^{10}$ M$_{\odot}$ and a transition in orientation at an HI mass of $\sim10^{9.5}$ M$_{\odot}$, both based on the spin of the stellar component. Observationally, \citet{Welker2019} find a stellar transition mass from aligned to perpendicular orientations of galaxy spins bracketed between $10^{10.4}$ M$_{\odot}$, and $10^{10.9}$ M$_{\odot}$. They also compare their results to simulations finding that the transition mass varies with the mass scale used to define the filaments and that more refined filaments seem to lead to lower transition masses. Figure 15 shows the spin-filament angles as a function of mass. If the three smallest galaxies are removed (1213496, 1180660 $\&$ 1227948), there seems to be a hint of transition in orientation just around $\sim10^{10}$ M$_{\odot}$ in stellar mass (Figure 15a) and a more bracketed hint of transition just below $\sim10^{10}$ M$_{\odot}$ in HI mass (Figure 15b), both based on the stellar spins. At most, these are marginal hints of the predicted trends. Figure 15c and 15d show a less convincing trend in transition for the HI spin in relation to stellar mass and HI mass. The results from our small sample are quite interesting and provide an opportunity to look at individual galaxies. However, a larger sample with better statistics is needed to draw firm conclusions and provide more profound insight into the alignment of spin vectors with filaments.

\begin{figure*}
\includegraphics[scale=1.1]{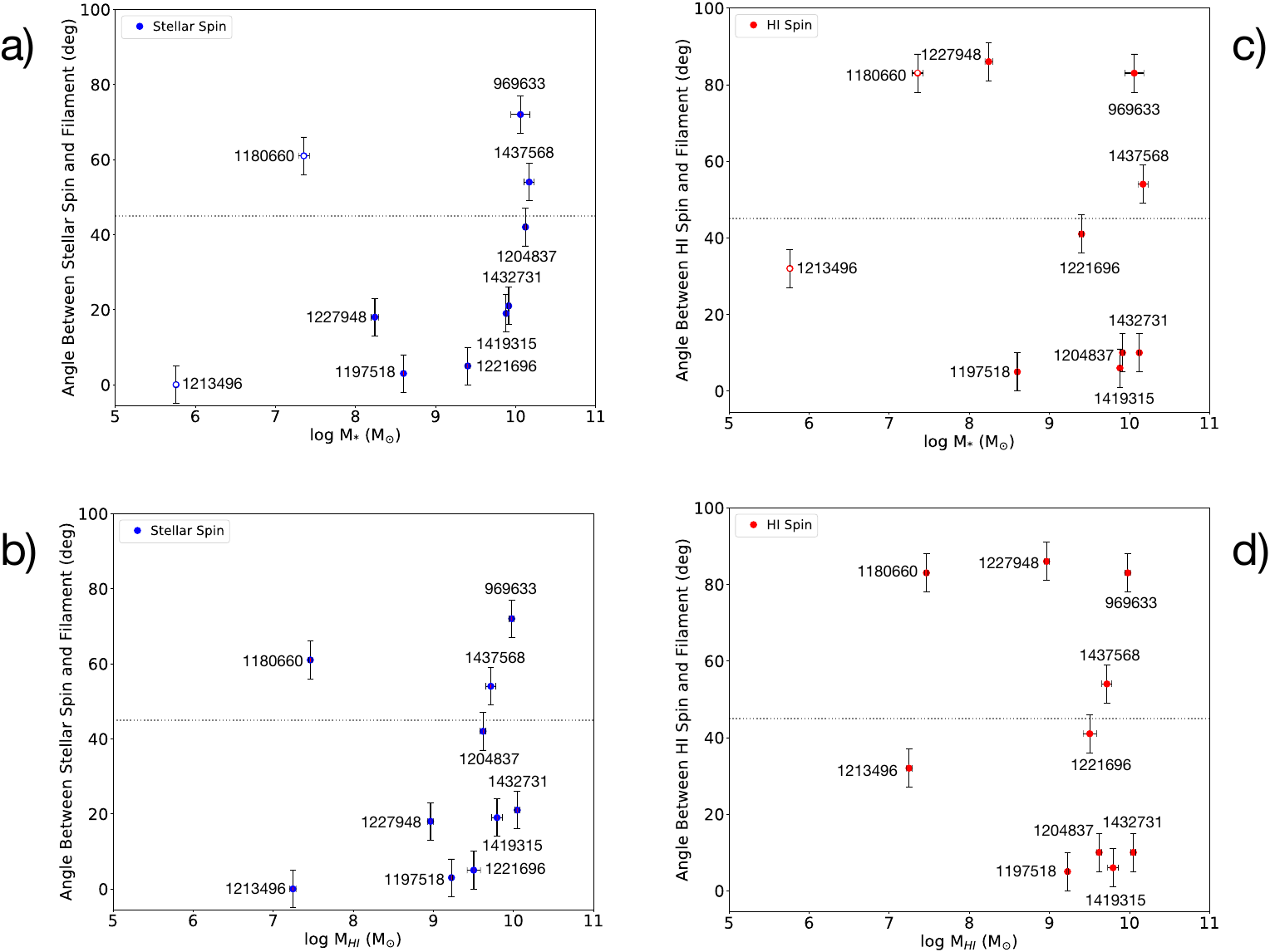}
\caption{Relation of the spin-filament angle to mass. Figure 15a shows the angle between the stellar spin and filament vs. stellar mass. Figure 15b shows the angle between the stellar spin and filament vs. HI mass. Figure 15c shows the angle between the HI spin and filament vs. stellar mass. Figure 15d shows the angle between the HI spin and filament vs. HI mass. A dotted grey line is drawn at 45 degrees. Open symbols are small galaxies with uncertain stellar masses.}
\label{figure15}
\end{figure*}

\section{Conclusions}
We present HI observations of ten galaxies out to a redshift of 0.1 from the first epoch of CHILES. We find a large fraction (nine out of ten) of galaxies is somewhat disturbed, with large HI disks measuring the flat part of the rotation curve. The two dwarf irregular galaxies have disturbances in their outer disks that are indicative of polar rings and may be an indication that these two galaxies are gravitationally interacting. We compare the neutral and ionized gas for our sample of galaxies. It appears that H$\alpha$ rotation curves approach the flat part, making them useful for measuring dark matter content and TF studies.

We explore galaxy properties as a function of location to cosmic web filaments. We find that galaxy spins tend to be aligned with cosmic web filaments and a hint of the predicted transition mass associated with the spin angle alignment. Our sample of galaxies lies within a range of distances from the cosmic web with the HI mass increases with increasing distance from filaments and SSFR decreases with increasing distance from filaments.

We utilize this small sample of nearby galaxies for a science verification study using our first 178 hours. We find that our data follow known scaling relations and demonstrate the use of cosmic web filaments to study the impact of the large-scale environment on these systems. With the full 1000 hours of the CHILES survey, we will be able to further study galaxy properties and orientation with respect to the cosmic web with images of close to 400 galaxies, using a broader mass sample, over the continuous redshift range (0 $<$ z $<$ 0.45) of the entire survey.

\section{Acknowledgements}
We would like to thank reviewer E. Tempel for very useful and constructive comments. We also thank K. Kraljic for very helpful comments on the paper. We would like to thank the entire CHILES collaboration. We additionally thank Steve Crawford and Ralf Kotulla for their assistance with SALT data reduction. We would like to also thank K. Vinsen and E. da Cunha who ran the SED fitting on the G10/COSMOS v05 catalog with MAGPHYS. J. Davis acknowledges support by the National Science Foundation (NSF) Graduate Research Fellowship Program under grant No. DGE-1747503 and NASA under Award No. NNX15AJ12H issued through Wisconsin Space Grant Consortium. D.J. Pisano and N. Luber acknowledge partial support from NSF grant No. AST 1412578. J.M.v.d.H. acknowledges support from the European Research Council under the European Union's Seventh Framework Programme (FP/2007-2013)/ERC Grant Agreement nr. 291531. M. Yun and H. B. Gim acknowledge support from NSF grant No. AST 1412843. This work is in part supported by the NSF under grant No. AST 141302 to Columbia University. Support for this work is also provided by the NSF through award SOSP 18$\_$3133 from the National Radio Astronomy Observatory (NRAO). The NRAO is a facility of the NSF operated under cooperative agreement by Associated Universities, Inc.

This research made use of Astropy, a community-developed core Python package for Astronomy (Astropy Collaboration 2013). Rotation curve overlays are made using Matplotlib (Hunter 2007).\\
Facilities: NED, SALT, VLA\\
Software: Astropy, CASA, IDL, IRAF, PySALT 

\bibliographystyle{mnras}
%\bibliography{references.bib}

\appendix
\section{Figures for Individual Galaxies}
We present figures for individual galaxies (Figures A1 - A10).

\begin{figure*}
\includegraphics[scale=0.62]{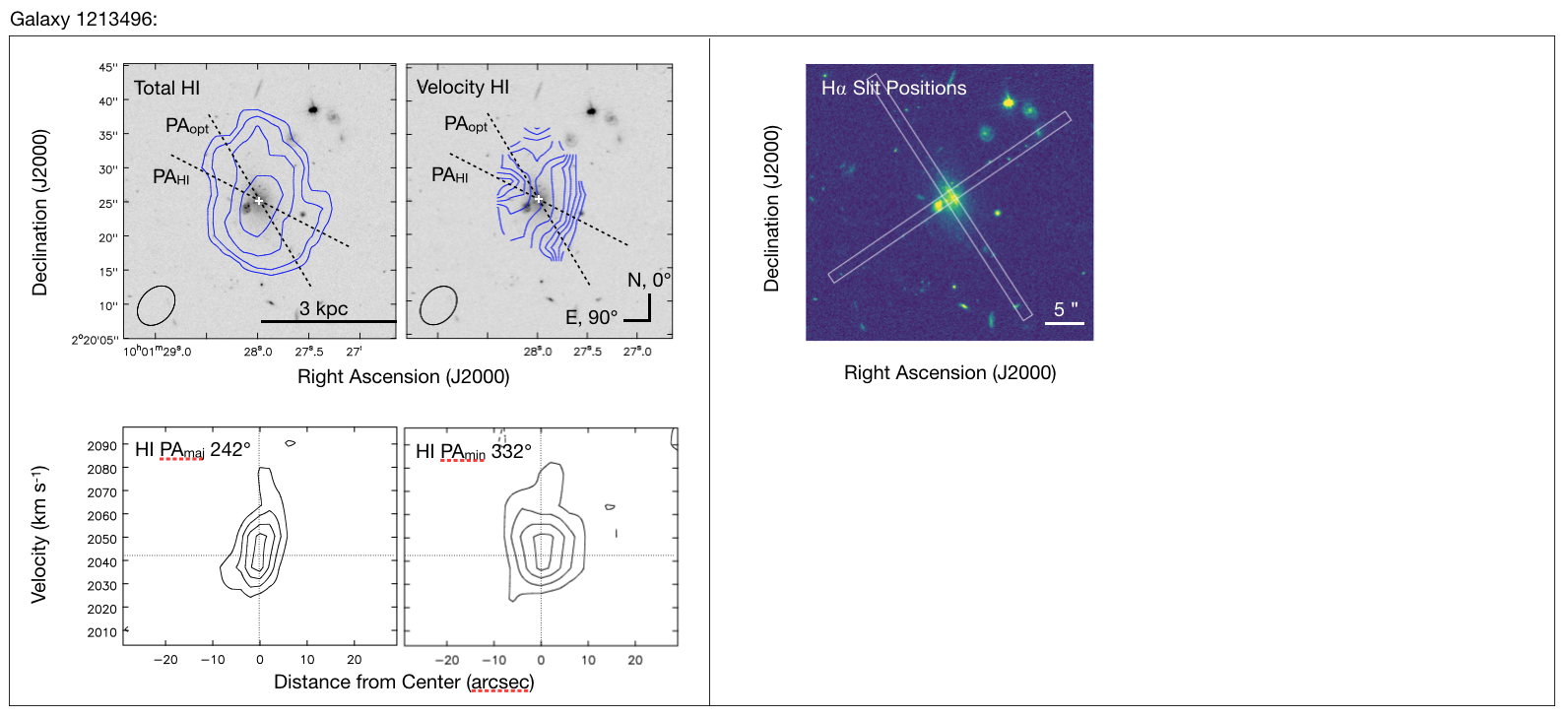}
\caption{Properties for galaxy 1213496. Detailed descriptions can be found in Section 3.6. HI PV diagrams$\colon$ 76 (RMS) $\times$ -4, -2 (dashed), 2, 4, 6, 8 $\mu Jy$ beam$^{-1}$; HI Total Integrated Flux$\colon$ 6.6 (2$\sigma$, 13.3 km s$^{-1}$ channel), 13.2, 26.5, 53.0 $\times$ 10$^{19}$ cm$^{-2}$; HI Velocity Field$\colon$ 2041 (system) $\pm$ 5 km s$^{-1}$; HI PA$\colon$ HI major axis PA is taken from the receding side. The global HI profile is shown in Figure 3. No H$\alpha$ is detected for this galaxy.}
\end{figure*}

\begin{figure*}
\includegraphics[scale=0.62]{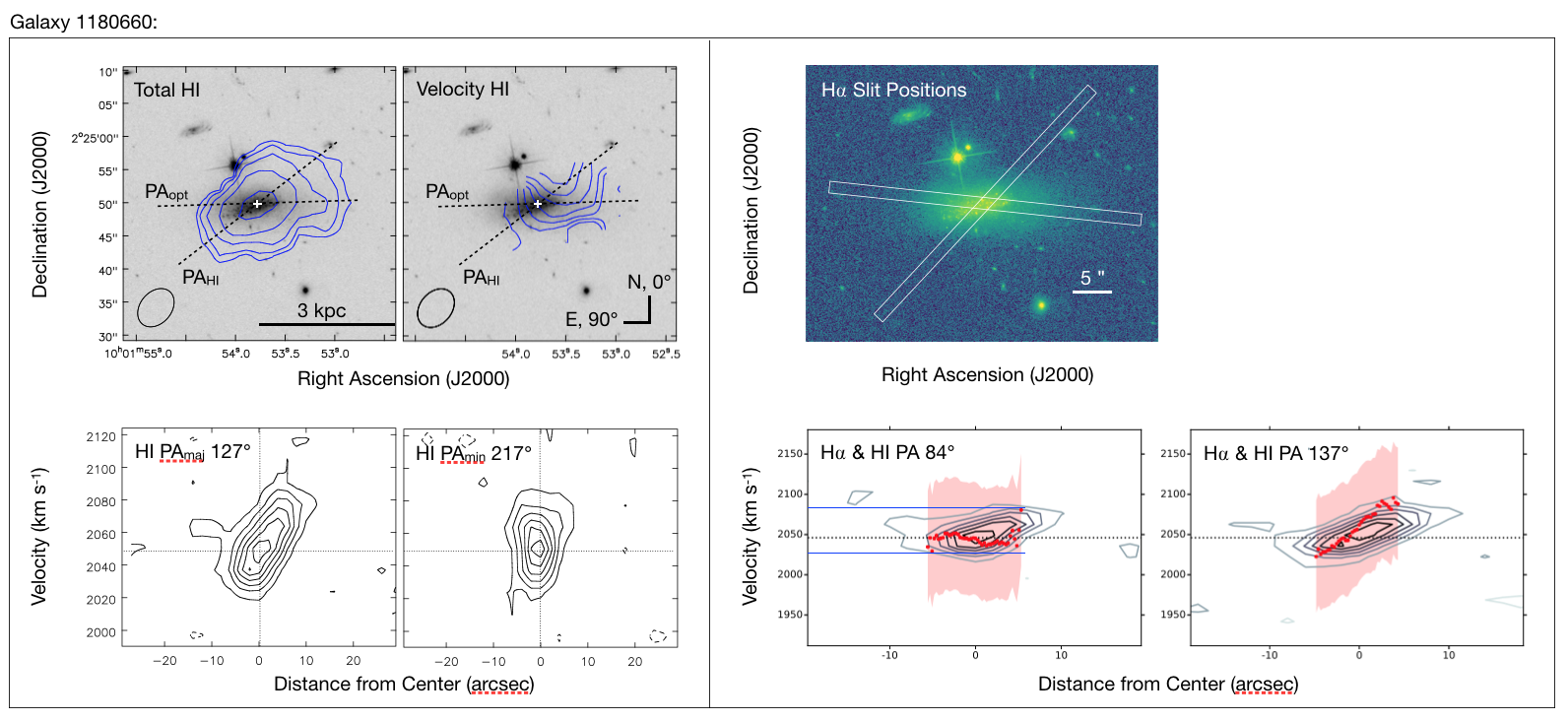}
\caption{Properties for galaxy 1180660. Detailed descriptions can be found in Section 3.6. HI PV diagrams$\colon$ 76 (RMS) $\times$ -4, -2 (dashed), 2, 4, 6, 8, 10, 12 $\mu Jy$  beam$^{-1}$; HI Total Integrated Flux$\colon$ 6.6 (2$\sigma$, 13.3 km s$^{-1}$ channel), 13.2, 26.5, 53.0, 106.0 $\times$ 10$^{19}$ cm$^{-2}$; HI Velocity Field$\colon$ 2046 (system) $\pm$ 3 km s$^{-1}$; HI PA$\colon$ HI major axis PA is taken from the receding side. The global HI profile is shown in Figure 3. H$\alpha$ is represented by the red points, and the red shaded region is the SALT velocity resolution ($\sim$ 70 km s$^{-1}$), taken here as the uncertainty.}
\end{figure*}

\begin{figure*}
\includegraphics[scale=0.61]{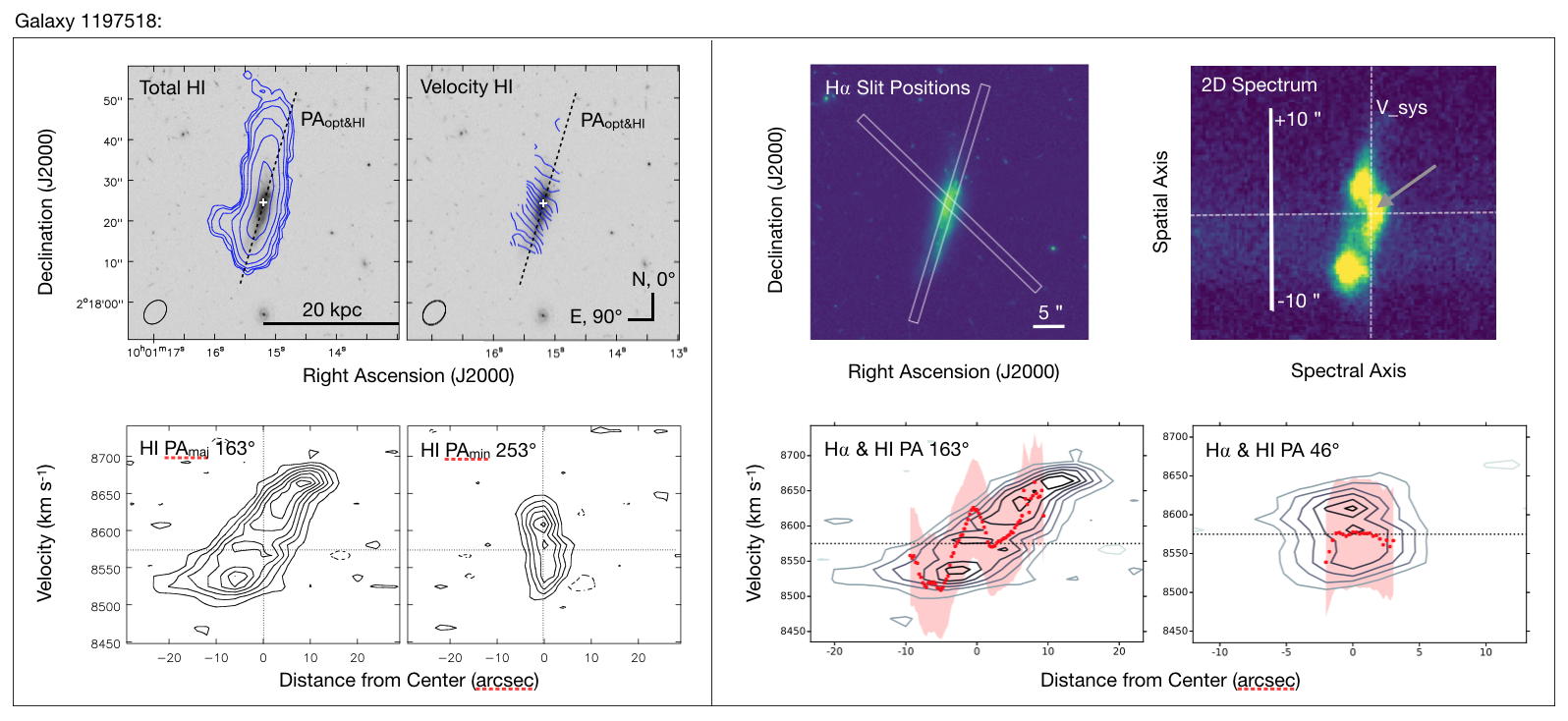}
\caption{Properties for galaxy 1197518. Detailed descriptions can be found in Section 3.6. HI PV diagrams$\colon$ 74 (RMS) $\times$ -4, -2 (dashed), 2, 4, 6, 8, 10, 12, 14 $\mu Jy$  beam$^{-1}$; HI Total Integrated Flux$\colon$ 6.8 (2$\sigma$, 13.6 km s$^{-1}$ channel), 13.6, 27.3, 54.6, 109.0 $\times$ 10$^{19}$ cm$^{-2}$; HI Velocity Field$\colon$ 8575 (system) $\pm$ 10 km s$^{-1}$; HI PA$\colon$ HI major axis PA is taken from the receding side. The global HI profile is shown in Figure 3. H$\alpha$ is represented by the red points, and the red shaded region is the SALT velocity resolution ($\sim$ 70 km s$^{-1}$), taken here as the uncertainty. The upper rightmost image is the 2D SALT spectrum, showing the H$\alpha$ emission line. The emission line is poorly fit by a single Gaussian and results in the H$\alpha$ rotation curve kink, due to the clump indicated by the grey arrow on the redward side of the normal rotation component. This clump could also be gas in non-circular motion due to the bar, and in front so we do not see the corresponding back side due to extinction.}
\end{figure*}

\begin{figure*}
\includegraphics[scale=0.61]{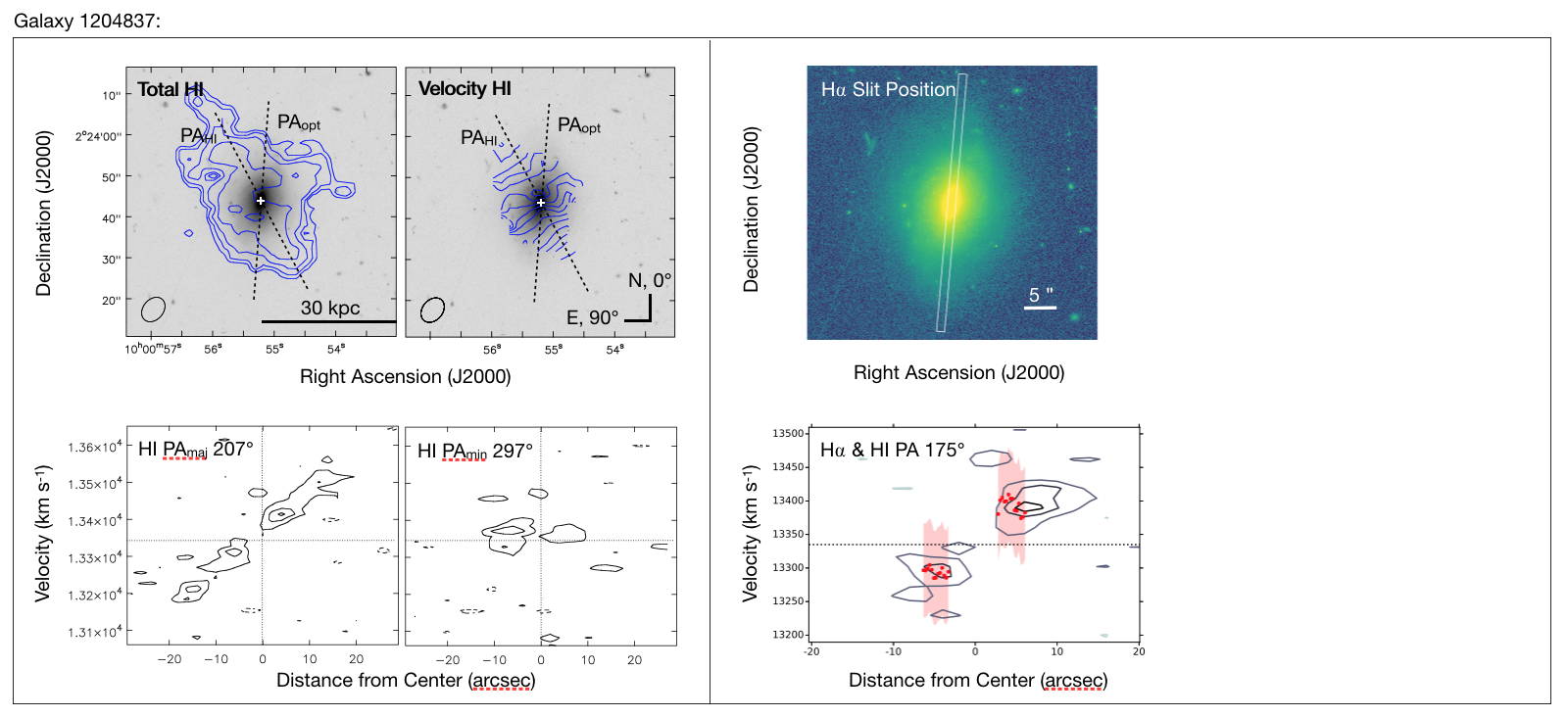}
\caption{Properties for galaxy 1204837. Detailed descriptions can be found in Section 3.6. HI PV diagrams$\colon$ 79 (RMS) $\times$ -4, -2 (dashed), 2, 4, 6 $\mu Jy$  beam$^{-1}$; HI Total Integrated Flux$\colon$ 7.5 (2$\sigma$, 13.8 km s$^{-1}$ channel), 14.9, 29.8, 59.7, 119.0 $\times$ 10$^{19}$ cm$^{-2}$; HI Velocity Field$\colon$ 13335 (system) $\pm$ 20 km s$^{-1}$; HI PA$\colon$ HI major axis PA is taken from the receding side. The global HI profile is shown in Figure 3. H$\alpha$ is represented by the red points, and the red shaded region is the SALT velocity resolution ($\sim$ 70 km s$^{-1}$), taken here as the uncertainty.}
\end{figure*}

\begin{figure*}
\includegraphics[scale=0.61]{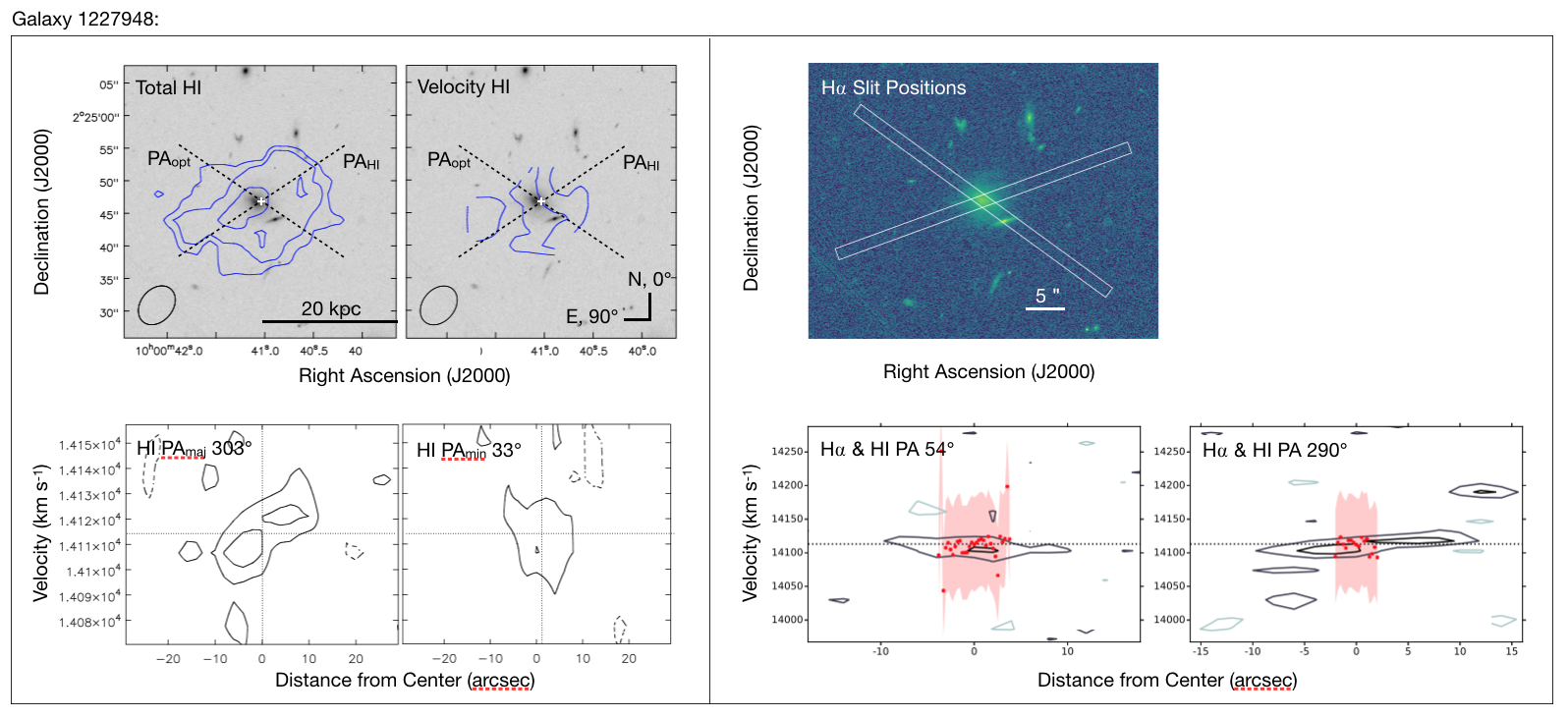}
\caption{Properties for galaxy 1227948. Detailed descriptions can be found in Section 3.6. HI PV diagrams$\colon$ 79 (RMS) $\times$ -4, -2 (dashed), 2, 4 $\mu Jy$  beam$^{-1}$; HI Total Integrated Flux$\colon$ 7.5 (2$\sigma$, 13.8 km s$^{-1}$ channel), 15.1, 30.1 $\times$ 10$^{19}$ cm$^{-2}$; HI Velocity Field$\colon$ 14113 (system) $\pm$ 5 km s$^{-1}$; HI PA$\colon$ HI major axis PA is taken from the receding side. The global HI profile is shown in Figure 3. H$\alpha$ is represented by the red points, and the red shaded region is the SALT velocity resolution ($\sim$ 70 km s$^{-1}$), taken here as the uncertainty.}
\end{figure*}

\begin{figure*}
\includegraphics[scale=0.61]{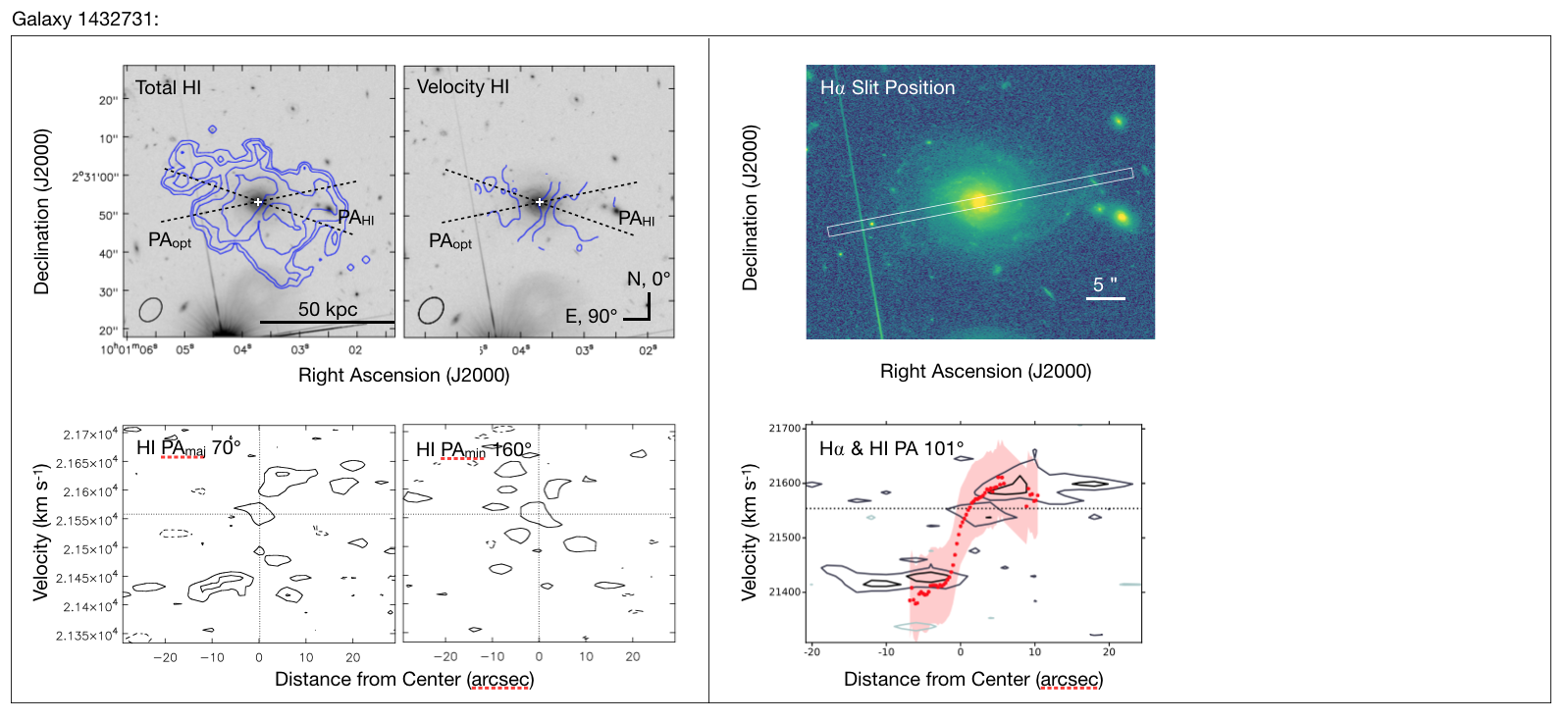}
\caption{Properties for galaxy 1432731. Detailed descriptions can be found in Section 3.6. HI PV diagrams$\colon$ 86 (RMS) $\times$ -4, -2 (dashed), 2, 4 $\mu Jy$  beam$^{-1}$; HI Total Integrated Flux$\colon$ 8.5 (2$\sigma$, 14.1 km s$^{-1}$ channel), 17.0, 34.1, 68.2 $\times$ 10$^{19}$ cm$^{-2}$; HI Velocity Field$\colon$  21554 (system) $\pm$ 50 km s$^{-1}$; HI PA$\colon$ HI major axis PA is taken from the receding side. The global HI profile is shown in Figure 3. H$\alpha$ is represented by the red points, and the red shaded region is the SALT velocity resolution ($\sim$ 70 km s$^{-1}$), taken here as the uncertainty. Note: The system velocity V$_{sys}$ is indicated with the horizontal dotted line in the PV diagrams and is taken as the velocity value at the optical center of the velocity field. For G146, this value appears higher than where the system velocity is expected to fall on the PV diagrams, at the center between the maximum velocity of the rising and declining parts of the HI emission.}
\end{figure*}

\begin{figure*}
\includegraphics[scale=0.61]{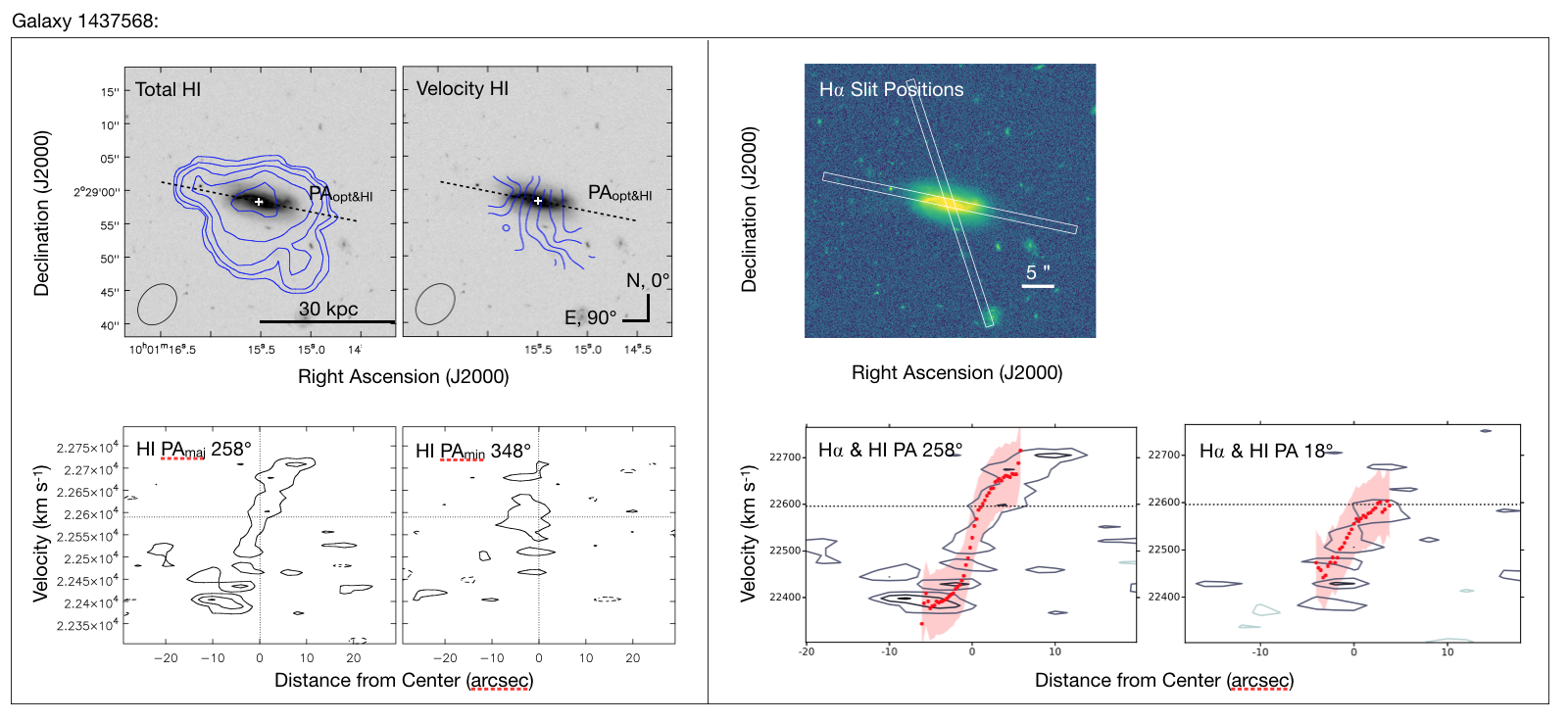}
\caption{Properties for galaxy 1437568. Detailed descriptions can be found in Section 3.6. HI PV diagrams$\colon$ 86 (RMS) $\times$ -4, -2 (dashed), 2, 4, 6 $\mu Jy$  beam$^{-1}$; HI Total Integrated Flux$\colon$  8.6 (2$\sigma$, 14.1 km s$^{-1}$ channel), 17.3, 34.6, 69.1, 138 .0 $\times$ 10$^{19}$ cm$^{-2}$; HI Velocity Field$\colon$ 22596 (system) $\pm$ 30 km s$^{-1}$; HI PA$\colon$ HI major axis PA is taken from the receding side. The global HI profile is shown in Figure 3. H$\alpha$ is represented by the red points, and the red shaded region is the SALT velocity resolution ($\sim$ 70 km s$^{-1}$), taken here as the uncertainty. Note: The system velocity V$_{sys}$ is indicated with the horizontal dotted line in the PV diagrams and is taken as the velocity value at the optical center of the velocity field. For 1437568, this value appears higher than where the system velocity is expected to fall on the PV diagrams, at the center between the maximum velocity of the rising and declining parts of the HI emission.}
\end{figure*}

\begin{figure*}
\includegraphics[scale=0.61]{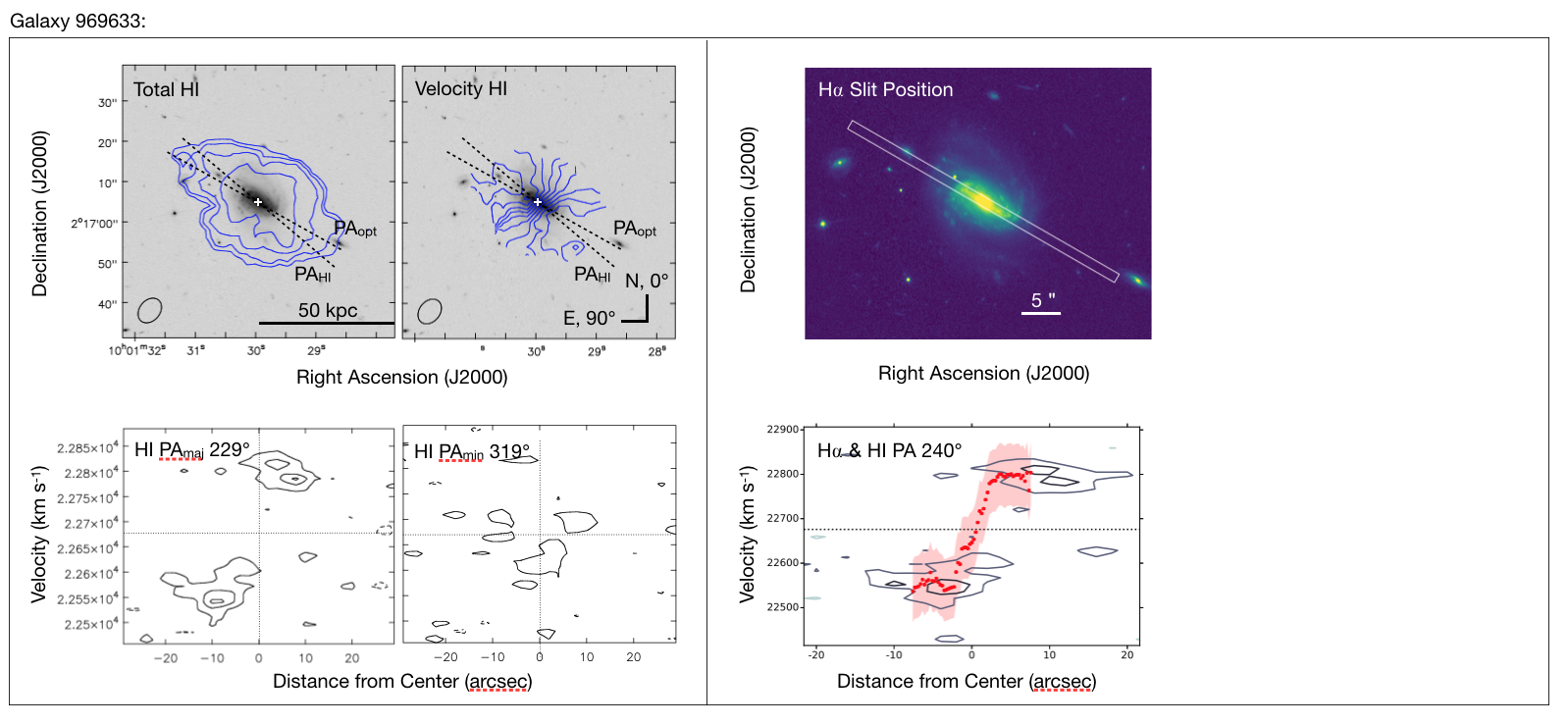}
\caption{Properties for galaxy 969633. Detailed descriptions can be found in Section 3.6. HI PV diagrams$\colon$ 86 (RMS) $\times$ -4, -2 (dashed), 2, 4, 6 $\mu Jy$  beam$^{-1}$; HI Total Integrated Flux$\colon$  8.6 (2$\sigma$, 14.1 km s$^{-1}$ channel), 17.3, 34.6, 69.1 $\times$ 10$^{19}$ cm$^{-2}$; HI Velocity Field$\colon$  22676 (system) $\pm$ 25 km s$^{-1}$; HI PA$\colon$ HI major axis PA is taken from the receding side. The global HI profile is shown in Figure 3. H$\alpha$ is represented by the red points, and the red shaded region is the SALT velocity resolution ($\sim$ 70 km s$^{-1}$), taken here as the uncertainty.}
\end{figure*}

\begin{figure*}
\includegraphics[scale=0.61]{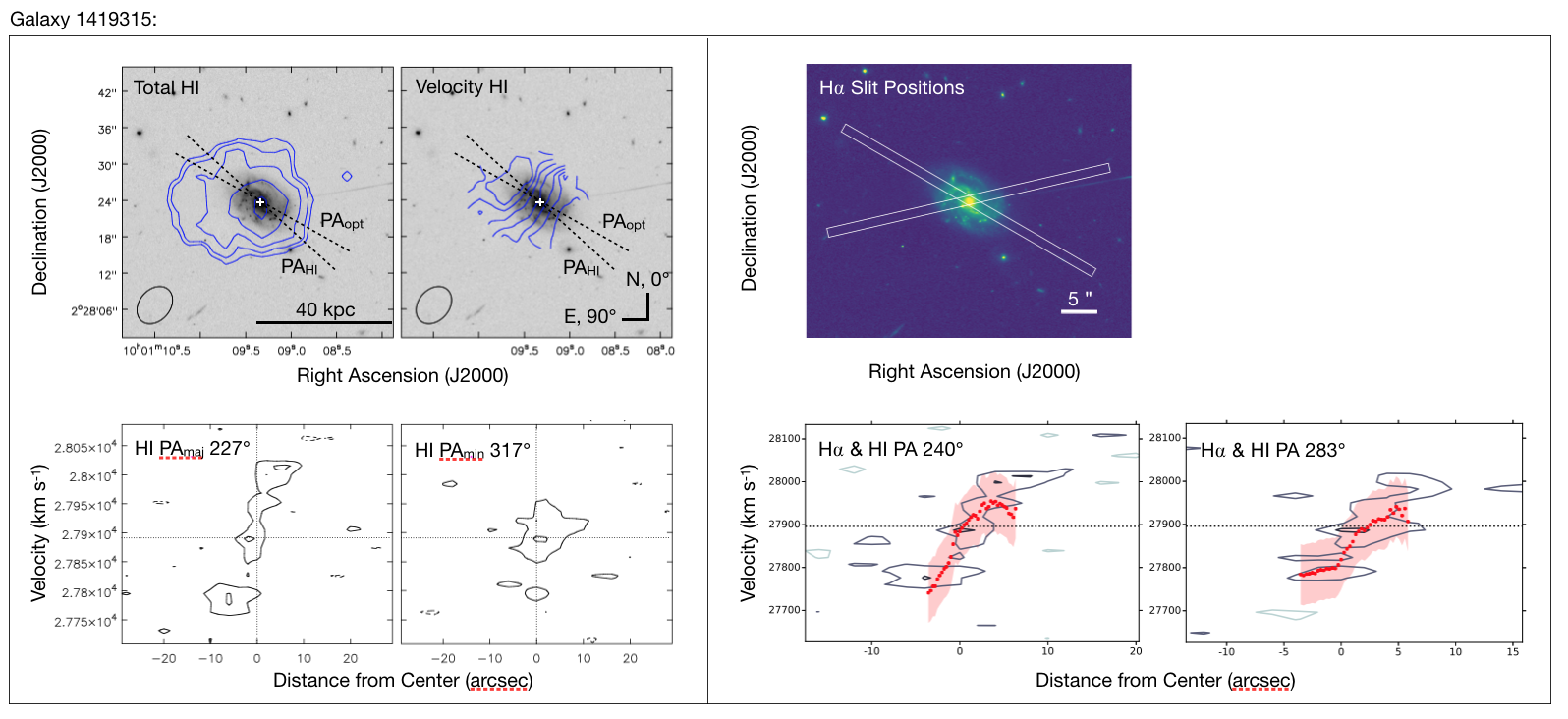}
\caption{Properties for galaxy 1419315. Detailed descriptions can be found in Section 3.6. HI PV diagrams$\colon$ 83 (RMS) $\times$ -4, -2 (dashed), 2, 4 $\mu Jy$  beam$^{-1}$; HI Total Integrated Flux$\colon$ 8.8 (2$\sigma$, 14.3 km s$^{-1}$ channel), 17.6, 35.3, 70.5 $\times$ 10$^{19}$ cm$^{-2}$; HI Velocity Field$\colon$ 27896 (system) $\pm$ 30 km s$^{-1}$; HI PA$\colon$ HI major axis PA is taken from the receding side. The global HI profile is shown in Figure 3. H$\alpha$ is represented by the red points, and the red shaded region is the SALT velocity resolution ($\sim$ 70 km s$^{-1}$), taken here as the uncertainty.}
\end{figure*}

\begin{figure*}
\includegraphics[scale=0.61]{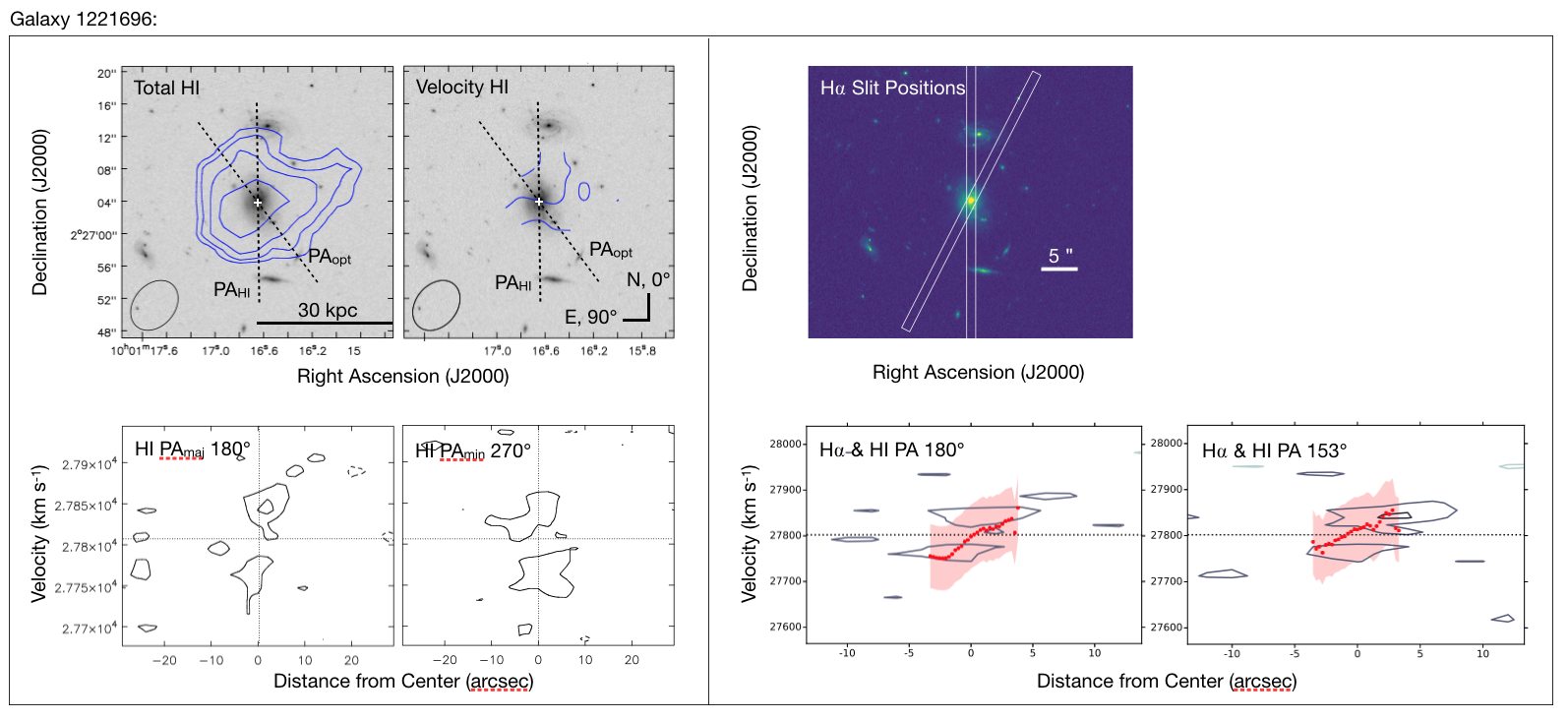}
\caption{Properties for galaxy 1221696. Detailed descriptions can be found in Section 3.6. PV diagrams$\colon$ 83 (RMS) $\times$ -4, -2 (dashed), 2, 4 $\mu Jy$  beam$^{-1}$; HI Total Integrated Flux$\colon$ 8.8 (2$\sigma$, 14.3 km s$^{-1}$ channel), 17.6, 35.2, 70.5 $\times$ 10$^{19}$ cm$^{-2}$; HI Velocity Field$\colon$ 27802 (system) $\pm$ 30 km s$^{-1}$; HI PA$\colon$ HI major axis PA is taken from the receding side. The global HI profile is shown in Figure 3. H$\alpha$ is represented by the red points, and the red shaded region is the SALT velocity resolution ($\sim$ 70 km s$^{-1}$), taken here as the uncertainty. }
\end{figure*}

% \bsp	
\label{lastpage}
\end{document}